\documentstyle[prb,aps,psfig]{revtex}

\begin{document}

\title{  A numerical study of multi-soliton configurations in a doped
antiferromagnetic Mott insulator}
\author{  Mona Berciu and Sajeev John}
\address{
Department of Physics, University of Toronto,
60 St. George Street, Toronto, Ontario, M5S 1A7, Canada}
\date{\today}
\maketitle
\pacs{Nos. 74.20.Mn, 74.25.Ha, 71.27.+a }

\begin{abstract}
We evaluate from first principles the self-consistent Hartree-Fock
energies for multi-soliton configurations in a doped, spin-${1 \over
2}$, antiferromagnetic Mott insulator on a two-dimensional square
lattice. The microscopic Hamiltonian for this system involves a
nearest neighbor electron hopping matrix element $t$, an on-site
Coulomb repulsion $U$, and a nearest neighbor Coulomb repulsion
$V$. We find that nearest-neighbor Coulomb repulsion on the energy
scale of $t$ stabilizes a regime of charged meron-antimeron vortex
soliton pairs over a region of doping from $\delta=0.05$ to $ 0.4$
holes per site for intermediate coupling $3 \le U/t \le 8$.  This
stabilization is mediated through the generation of ``spin-flux'' in
the mean-field antiferromagnetic (AFM) background.  Spin-flux is a new
form of spontaneous symmetry breaking in a strongly correlated
electron system in which the Hamiltonian acquires a term with the
symmetry of spin-orbit coupling at the mean-field level.  Spin-flux
modifies the single quasi-particle dispersion relations from that of a
conventional AFM. The modified dispersion is consistent with
angle-resolved photo-emission studies and has a local minimum at
wavevector $\vec{k}=\pi/2a(1,1)$, where $a$ is the lattice constant.
Holes cloaked by a meron-vortex in the spin-flux AFM background are
charged bosons. Our static Hartree-Fock calculations provide an upper
bound on the energy of a finite density of charged vortices. This
upper bound is lower than the energy of the corresponding charged
spin-polaron configurations. A finite density of charge carrying
vortices is shown to produce a large number of unoccupied electronic
levels in the Mott-Hubbard charge transfer gap. These levels lead to
significant band tailing and a broad mid-infrared band in the optical
absorption spectrum as observed experimentally.  In the presence of a
finite density of charged meron-antimeron pairs, the peak in the
magnetic structure at $\vec{Q}=\pi/a (1,1)$, corresponding to the
undoped AFM, splits into four satellite peaks which evolve with charge
carrier concentration as observed experimentally.  At very low doping
($\delta < 0.05)$ the doping charges create extremely tightly bound
meron-antimeron pairs or even isolated conventional spin-polarons,
whereas for very high doping $(\delta > 0.4)$ the spin background
itself becomes unstable to formation of a conventional Fermi liquid
and the spin-flux mean-field is energetically unfavorable.  Our
results point to the predominance of a quantum liquid of charged,
bosonic, vortex solitons at intermediate coupling and intermediate
doping concentrations.
\end{abstract}

\narrowtext
\twocolumn

\section{Introduction}

A microscopic description of doped, spin-${1\over 2}$, Mott insulators
\cite{a,b} is a central issue in the understanding of strongly
correlated electrons and high-temperature superconductivity. \cite{c}
Of particular interest is the intermediate doping regime of $\delta
=0.05-0.30$ charge carriers per lattice site on a two-dimensional
square lattice. In this regime it has been observed, in a variety of
cuprate superconducting materials, that long range antiferromagnetic
(AFM) order is destroyed by the presence of charge carriers and that
these charge carriers lead to a variety of non-Fermi-liquid
characteristics in the transport and electromagnetic response. The
electrical, \cite{d} magnetic \cite{e} and optical properties
\cite{f} of the doped parent compound, from which superconductivity
emerges, are among the most glaring and profound mysteries in solid
state physics today. \cite{g,h}

	In most theoretical studies of the doped Mott insulator, it
has been assumed that the Coulomb repulsion between electrons can be
described by a Hubbard model in which this interaction is replaced by
a point-like on-site interaction. Moreover, it has been assumed that
the on-site Hubbard parameter $U$ is an order of magnitude larger than
the nearest neighbor electron hopping energy scale, $t$. This picture
is based on the Fermi-liquid theory notion of screening of the
effective electron-electron interaction. However, it is this same
Fermi liquid picture which studies of the Hubbard model seek to
supplant. In some recent papers \cite{1,2,3,3b} we have shown that the
nearest neighbor Coulomb interaction is a highly relevant perturbation
which can lead to an entirely new type of broken symmetry in the
many-electron system. At the mean-field level we showed that Coulomb
effects may give rise to a mean-field (Hartree-Fock) state in which
the Hamiltonian acquires a term with the symmetry of a spin-orbit
interaction. In this state, which we refer to as a spin-flux phase,
the internal wavefunction of the electron (in spin-space) undergoes a
$2\pi$-rotation as the electron encircles any elementary plaquette of
the 2D square lattice.  It was shown that spin-flux itself is a
dynamical variable which appears in quantized units and is carried by
neutral skyrmion textures even within the undoped AFM.  It was shown
\cite{2} that in the presence of a mean-field of such spin-flux, the
mean-field ground-state energy is lower than in the absence of the
spin-flux (i.e. in the conventional AFM phase), for a large range of
doping concentrations and on-site repulsion strength $U$.

These early comparisons of the spin-flux states with non-spin-flux
states \cite{2} assumed that the doping electrons (holes) formed
extended states within the Mott-Hubbard bands leading to a global
twist of the AFM background into a single wave-vector, incommensurate
spiral state. More recently we have shown that the added charge
carriers find it energetically favorable, in the majority of cases, to
nucleate a point defect in the AFM background (charged magnetic
soliton) rather than occupy a band state. \cite{3,3b} The existence of
these charged solitons provides a remarkable and clear, microscopic
mechanism for non-Fermi-liquid behavior in a doped Mott insulator.
Starting from the microscopic many-electron Hamiltonian, we derived a
simple continuum model for the description of the magnetic soliton
textures, such as skyrmions and merons, which are generated by doping
of the antiferromagnetic parent compound.  This continuum model
recaptures the Mott-Hubbard gap structure by retaining the exact
electron dynamics and AFM spin correlations on the scale of the
elementary lattice plaquette. It is approximate in the sense that it
assumes that the local magnetic structure varies slowly from one
plaquette to the next and that the electron dispersion relations are
linearized about the relevant Fermi points.  The existence of magnetic
textures leads to the appearance of bound levels deep inside the
Mott-Hubbard charge-transfer gap. A doping hole can considerably lower
its energy by occupying such a bound level, with the effect that the
hole becomes trapped in the core of the magnetic soliton, in turn
stabilizing the soliton. This leads to a striking analogy between the
2D AFM and the 1D polyacetylene.\cite{4} In both cases, the mean-field
ground-state of the undoped system is degenerate as a result of broken
symmetry. In both cases, doping induces topological fluctuations
(solitons) that tend to restore the broken symmetry.  In particular,
we have shown that skyrmions (magnetic spin polarons) are the 2D
analogues of 1D polarons in polyacetylene, while charged
meron-vortices are the 2D analogues of the charged bosonic domain-wall
solitons in polyacetylene. \cite{5} At the topological level a polaron
in 1D can be thought of as a tightly bound pair of domain-walls,
whereas a skyrmion in 2D is topologically equivalent to a bound
meron-antimeron pair.  \cite{5bis} The analogy also holds at the level
of the electronic structure. Both the 1D domain-wall in polyacetylene
and the 2D charged meron-vortex in the AFM lead to the occurrence of
mid-gap electron states in their respective one-electron band
structures. It is well-known that in polyacetylene, the first charge
carrier added to the undoped polymer creates a polaron around itself,
while a second charge carrier causes this polaron to split into two
independent domain-walls, each carrying one dopant. We suggested
\cite{3b} that a similar picture holds in the 2D AFM: the first hole
is cloaked by a magnetic spin-polaron, while a second hole causes the
polaron to split into a bound meron-antimeron pair, each vortex
carrying a doping charge.

Neutral vortex-antivortex bound pairs may appear without doping in the
layered AFM parent compound as the temperature is increased. \cite{6}
At finite doping, similar charged pairs appear even at $T=0$, since
the increase in energy due to the distortion in the AFM background is
compensated by the energy gained through trapping the holes in mid-gap
states near the vortex-cores.  If the doping reaches a critical value,
these pairs may unbind even at $T=0$ leading to the destruction of
long range AFM order.  For dopings smaller than this critical doping,
a finite temperature may also entropically drive the transition from
the AFM ordered state to a disordered ``spin-liquid''.

	In this article we treat the $T=0$ case, and demonstrate that
a transition from a dilute gas of charged spin-polarons to a liquid of
charged meron-vortex solitons takes place for intermediate doping and
for intermediate values of $U/t$, in the Hartree-Fock picture.  The
meron-vortices are bosonic charge carriers, with deep electronic gap
levels localized in their cores. The bosonic character may provide an
explanation for the unusual non-Fermi-liquid properties of the metal
observed in the intermediate doping region, while the deep gap
electronic structure may be related to the doping-induced band-tailing
effects and the observed broad mid-infrared optical absorption
band. At higher dopings, we show that the conventional phase with
fermionic charge carriers has a lower Hartree-Fock energy. This is
consistent with the observed transition to a normal metal when the
cuprates superconductors are overdoped.

Consider a strongly interacting quasi-two-dimensional electron gas
described by the tight-binding Hamiltonian
\begin{equation}
\label{1.1}
{\cal H} = -\sum_{{\langle ij \rangle \atop \sigma}}
t_{ij}(a_{i\sigma}^+a_{j\sigma} +
h.c.) + \sum_{ij}V_{ij}n_in_j 
\end{equation}
where $a_{i\sigma}^+$ creates an electron at site $i$ with spin
$\sigma$, $t_{ij}$ is the hopping amplitudes from site $j$ to site $i$
on the square lattice, $\hat{n}_i\equiv
\sum\limits^2_{\sigma=1}a_{i\sigma}^+a_{i\sigma}$, and $V_{ij}$ is the
Coulomb interaction.  For nearest neighbor hopping $(t_{ij} = t_0)$
and purely on-site Coulomb repulsion $(V_{ii}\equiv U)$, this reduces
to the one-band Hubbard model.  In order to describe the possibility
of spin rotation during the process of electron hopping, we retain the
{\it nearest neighbor} Coulomb repulsion $(V_{ij}=V)$.  Using the
Pauli spin-matrix identity, ${1\over2}
\sigma^\mu_{\alpha\beta}(\sigma^\mu_{\alpha^\prime\beta^\prime})^ \ast
=\delta_{\alpha\alpha^\prime} \delta_{\beta\beta^\prime}$, it is
possible to rewrite the electron-electron interaction terms in the
exact form: $n_in_j=(2+\delta_{ij})n_i -{1 \over 2}
\Lambda^\mu_{ij}(\Lambda_{ij}^\mu)^+$.  Here, $\Lambda^\mu_{ij}$ are
bilinear combinations of electron operators defined by
$\Lambda^\mu_{ij}\equiv
a^+_{i\alpha}\sigma^\mu_{\alpha\beta}a_{j\beta}$, $\mu=0, 1, 2,
3$. $\sigma^0$ is the $2\times 2$ identity matrix, $\vec \sigma\equiv
(\sigma^1, \sigma^2,\sigma^3)$ are the usual Pauli spin matrices and
there is implicit summation over the repeated indices.  The quantum
expectation value $\langle\ \rangle$ of the $\Lambda^{\mu}_{ij}$
operators, for $i \neq j$ are associated with charge-currents
($\mu=0$) and spin-currents ($\mu=1,2,3$).  Likewise, the quantum
expectation value of $\Lambda^{\mu}_{ij}$ for $i=j$ describes the
on-site charge density $Q_i= \Lambda^{0}_{ii}$ and the on-site
spin-density $S_i^a=\Lambda^{a}_{ii}$, $a=1,2,3$.  In the AFM
spin-flux model \cite{1,2,3,3b}, we adopt the ansatz that there are
no charge density waves (CDW) or charge currents in the ground state
$\Lambda^{0}_{ij}= 0$. For positive on-site Hubbard interaction, any
CDW would considerably increase the mean-field ground state
energy. Circulating charge currents are accompanied by magnetic fields
and have been considered in the context of \underline{conventional}
flux phases. \cite{7} However, such states are not observed
experimentally in the cuprate superconductors. On the other hand we
incorporate the experimentally observed AFM spin-density background
$\langle \vec{S}_i\rangle$, and we postulate the existence of
circulating ``spin-currents'' which take the form $\langle
\Lambda^{a}_{ij} \rangle ={ 2 t_o \over V} \Delta_{ij} \hat{n}_a$,
where $|\Delta_{ij}|=\Delta$ for all $(ij)$ and $\hat{n}$ is a unit
vector. In the spin-flux phase, these ``spin-currents'' do not cause
any rotation of the local magnetic moments $\langle
\vec{S}_i\rangle$. Instead, they correspond to rotations in the
internal space of Euler angles (phase changes) as the electrons
circulate around lattice plaquettes.

Implementing this ansatz with the help of the mean-field factorization
$\Lambda^\mu_{ij}(\Lambda^\mu_{ij})^+\rightarrow \langle
\Lambda^\mu_{ij}\rangle (\Lambda^\mu_{ij})^+ +\Lambda^\mu_{ij} \langle
\Lambda^\mu_{ij}\rangle^*+
\langle\Lambda^\mu_{ij}(\Lambda^\mu_{ij})^+\rangle
-2\langle\Lambda^\mu_{ij}\rangle \langle \Lambda^\mu_{ij}\rangle^*$, ,
for $i\neq j$ and making the Hartree-Fock factorization for the
on-site $i = j$ terms, we obtain the mean field Hamiltonian
\[ {\cal H}={\cal H}_{el}+{\cal H}_{const} \]
where
$$
{\cal H}_{el}=-t\sum_{\langle ij \rangle, \alpha \beta}^{} \left
( a^{\dagger}_{i,\alpha} T^{ij}_{\alpha\beta} a_{j,\beta}+
h.c. \right) - $$
\vspace{-5mm}
\begin{equation}
\label{1.4}
- U\sum_{i,\alpha,\beta}^{} a^{\dagger}_{i,\alpha} \left({\vec
S_i}\cdot{\vec \sigma}_{\alpha,\beta}\right) a_{i,\beta} +{U\over 2}
\sum_{i,\alpha}^{} \left(Q_i-1\right) a^{\dagger}_{i,\alpha}
a_{i,\alpha}
\end{equation}
and
\begin{equation}
\label{1.5}
{\cal H}_{const}=U\sum_{i}^{}\left(\vec{S}_i^2-{1\over 4}Q_i^2+{1\over
2} Q_i\right)
\end{equation}

\noindent Here, $T^{ij}_{\alpha\beta}\equiv (\delta_{\alpha\beta}
+i\Delta_{ij}\hat{n}\cdot\vec \sigma_{\alpha\beta})/\sqrt{1+\Delta^2}$
are spin-dependent $SU(2)$ hopping matrix elements defined by the
mean-field theory, and $t=t_o\sqrt{1+\Delta^2}$.

 In deriving (2) we have dropped constant terms as well as terms
proportional to $\sum\limits_in_i$ obtained from the mean-field
factorization of the nearest neighbor Coulomb interaction. However, we
have kept all terms obtained from the Hartree-Fock factorization of
the on-site Coulomb repulsion. Thus, the entire effect of the
nearest-neighbor Coulomb interaction is the renormalization of $t$ and
the appearance of the $T^{ij}_{\alpha\beta}$ phase-factors in the
hopping Hamiltonian.

	It was shown previously \cite{1,2} that the ground state
energy of the Hamiltonian of Eq. (2) depends on the SU(2) matrices
$T^{ij}$ only through the plaquette matrix product
$T^{12}T^{23}T^{34}T^{41}\equiv\exp (i\hat n\cdot\vec\sigma\Phi)$.
Here, $\Phi$ is the spin-flux which passes through each plaquette and
$2\Phi$ is the angle through which the internal coordinate system of
the electron rotates as it encircles the plaquette.  Since the
electron spinor wavefunction is two-valued, there are only two
possible choices for $\Phi$.  If $\Phi=0$ we can set
$T^{ij}_{\alpha\beta}=\delta_{ij}$ and the Hamiltonian (2) describes
conventional ordered magnetic states of the Hubbard model.  The other
possibility is that a spin-flux $\Phi=\pi$ penetrates each plaquette,
leading to $T^{12}T^{23}T^{34}T^{41}=-1$. This means that the
one-electron wavefunctions are antisymmetric around each of the
plaquettes, i.e. that as an electron encircles a plaquette, its
wavefunction in the internal spin space of Euler angles rotates by
$2\pi$ in response to strong interactions with the other electrons. We
call this the spin-flux phase. This uniform spin-flux phase is
accompanied by a AFM local moment background (with reduced magnitude)
and may be regarded as an alternative mean-field ground state of the
conventional AFM phase of the Hubbard model. In the spin-flux phase,
the kinetic energy term in (2) exhibits broken symmetry of a
spin-orbit type. This new form of spontaneous symmetry breaking occurs
over and above that associated with conventional
antiferromagnetism. It is also distinct from the smaller, conventional
spin-orbit effects which give rise to anisotropic corrections to
superexchange interactions between localized spins in the AFM. \cite{7bis}
We emphasize, however, that this AFM mean-field is a ``false ground
state''\cite{8} at finite doping, analogous to the ``false vacuum'' in
early models of quantum chromodynamics. \cite{9} In the presence of
charge carriers this mean-field is unstable to the proliferation of
topological fluctuations (magnetic solitons) which eventually destroy
AFM long range order. In this sense, the analysis which we present
below goes beyond simple mean field theory.

The article is organized as follows: in Section 2 we compare the
half-filled AFM mean-field ground states of the conventional phase and
the spin-flux phase. We show that the spin-flux phase mean-field
ground state always has a lower energy, and that it has a single
quasiparticle dispersion relation which is consistent with
angle-resolved photo-emission studies (ARPES).  This suggests that the
spin-flux phase is a suitable starting point for studying the behavior
of the parent compounds upon doping. In Section 3 we consider the
problem of adding just one hole to the AFM background. We study in
detail two possible soliton excitations, the fermionic, charged, spin
bag and the bosonic, charged, meron-vortex, for both the conventional
and the spin-flux phase. Using a simple energetical argument, we
propose a phase diagram for each of these excitations showing which is
the relevant excitation for various $U/t$ values and various
dopings. In the conventional phase we find that the spin bag is the
relevant excitation at all dopings and all values of $U/t$. In the
spin-flux phase, we find that for intermediate $U/t$ values and low
dopings, the meron-vortices are the relevant excitations. Since the
spin-flux phase has the lower energy, this means that a liquid of
meron-vortices appears on the lattice upon doping. This suggests a
plausible explanation for various unusual (non-Fermi-liquid)
properties of the underdoped and slightly overdoped cuprate
compounds. In Section 4 we study multi-soliton configurations, by
doping more holes into the lattice. The results obtained are in good
agreement with those predicted from the simple phase diagrams inferred
in Section 3. We also show that at higher doping (overdoped samples)
the conventional phase has a lower energy than the spin-flux phase,
and therefore a transition to a conventional Fermi-liquid takes place
in this regime. Using very simple assumptions, we calculate the
optical and static magnetic response of underdoped cuprate containing
a frozen liquid of meron vortices, and show that it is consistent with
the experimental measurements. Finally, Section 5 contains discussion
of the results and conclusions.

\section{The undoped Mott insulator}

In order to carry out Hartree-Fock calculations for multi-soliton
configurations in the antiferromagnet we consider a finite $N \times
N$ lattice. In this case the eigenvalues and eigenenergies of the
mean-field Hamiltonian can be found numerically,  and the convergence
algorithm is a straightforward iteration procedure.  Starting from an
initial spin and charge distribution, $\vec{S}(i)$ and $Q(i)$, for
$i=(i_x,i_y)$ with $i_x=1,N$, $i_y=1,N$, the mean-field Hamiltonian is
numerically diagonalized. This in turn leads to new expectation values
for the spin and charge distributions given by
$$
\vec{S}(i)= \sum_{\alpha=1}^{N_e} \sum_{\sigma, \sigma'=\pm 1}^{}
\phi_{\alpha}^*(i,\sigma)\left({1\over2}\vec{\sigma}
\right)_{\sigma\sigma'} \phi_{\alpha}(i,\sigma')
$$
$$ Q(i)=\sum_{\alpha=1}^{N_e} \sum_{\sigma=\pm
1}{}\phi_{\alpha}^*(i,\sigma)\phi_{\alpha}(i,\sigma)
$$
\noindent Here, $\alpha$ is an index for the eigenstates,
$\phi_{\alpha}$ is the corresponding eigenfunction, and $N_e$ is the
total number of electrons on the lattice. This is related to the
doping concentration (measured with respect to half-filling) by
$\delta=1-N_e/N^2$.  If the new spin and charge distribution are
different from the initial ones, we repeat the diagonalisation until
self-consistency is reached. In this article, self-consistency is
defined by the criterion that the largest variation of any of the
charge or spin components on any of the sites is less than $10^{-6}$
between successive iterations. We assume for simplicity that the
mean-field spin-flux parameters $T^{ij}$ are fixed. In a more general
theory, these may also be treated as dynamical variables.

It is experimentally observed that the ground-state of the undoped
Mott insulator has long-range AFM order. Accordingly, we choose a spin
distribution of the form $\vec{S}(i)= (-1)^{(i_x+i_y)}S \vec{e}$,
where $\vec{e}$ is the unit vector of some arbitrary direction, while
the charge distribution is $Q(i)=1$.  The results for the conventional
AFM are well known. In this case, we choose the Brillouin zone to be a
rotated square defined by $-{ \pi \over a} \le k_x + k_y, k_x - k_y
\le { \pi \over a}$. The dispersion relations are given by
\begin{equation}
\label{2.1} 
E^{\pm}(\vec{k})=\pm E(\vec{k})= \pm \sqrt{\epsilon^2(\vec{k})+(US)^2}
\end{equation}
where each level is two-fold degenerate and $\epsilon(\vec{k})
=-2t\left(\cos{(k_xa)}+\cos{(k_ya)}\right)$ is the one-electron
dispersion relation of the non-interacting conventional state. The
mean-field ground-state energy is given by (see Eqs. (\ref{1.4}),
(\ref{1.5}) )
\begin{equation}
\label{2.2} 
E^{gs}=-2\sum_{\vec{k}} E(\vec{k})+N^2 U\left(S^2+{1\over 4} \right).
\end{equation}
where the self-consistent value for the staggered spin $S$ satisfies
the energy minimization condition
\begin{equation}
\label{2.3}
S={1\over N^2}\sum_{\vec{k}}^{} {US \over E(\vec{k})}.
\end{equation}

In the spin-flux phase, it is more convenient to choose a square unit
cell, in order to simplify the description of the $T^{ij}$
phase-factors. We make the simplest gauge choice compatible with the
spin-flux condition for the $T$-matrices, namely that
$T^{12}=T^{23}=T^{34}=-T^{41}=1$ (see Fig. 1).  This leads to a
reduced square Brillouin zone $-\pi/2a \le k_x, k_y \le \pi/2a$.  The
dispersion relations for the AFM configuration are given by:
\begin{equation}
\label{2.4}
E^{\pm}_{sf}(\vec{k})=\pm E_{sf}(\vec{k})= \pm
\sqrt{\epsilon^2_{sf}(\vec{k})+(US)^2}
\end{equation}
where each level is four-fold degenerate and $\epsilon_{sf}(\vec{k})
=-2t \sqrt{\left(\cos{(k_xa)}\right)^2+\left( \cos{(k_ya)}\right)^2}$
are the noninteracting electron dispersion relations in the presence of
spin-flux.  The mean-field ground-state energy is given by
\begin{equation}
\label{2.5}
E^{gs}_{sf}=-4\sum_{\vec{k}} E_{sf}(\vec{k})+N^2 U\left(S^2+{1\over 4}
\right)
\end{equation}
where the AFM local moment amplitude is determined by the condition
\begin{equation}
\label{2.6}
S={2\over N^2}\sum_{\vec{k}}^{} {US \over E_{sf}(\vec{k})}.
\end{equation}

	In both the conventional and spin-flux phases, a Mott-Hubbard
gap of magnitude $2US$ opens between the valence and the conduction
bands. However, the Fermi surfaces are very different. In the
conventional phase, all the points of the Brillouin surface belong to
the nested Fermi surface, while in the spin-flux phase the Fermi
surface collapses to four points $\left ( \pm \pi/2a, \pm
\pi/2a\right)$. This means that the introduction of the spin-flux
leads to a lowering of the energies of all the other points of the
conventional nested Fermi surface, and thus for a strongly interacting
electron system the energy of the entire system is lower in the
spin-flux phase.  It is interesting to note that the quasi-particle
dispersion relation obtained in the presence of the spin-flux closely
resembles the dispersion as measured through angle-resolved
photo-emission studies (ARPES) in a compound such as
Sr$_2$CuO$_2$Cl$_2$ \cite{10}(see Fig. 2).  Namely, there is a a peak
centered at $(\pi/2, \pi/2)$ with an isotropic dispersion relation
around it, observed on both the $(0,0)$ to $(\pi,\pi)$ and $(0,\pi)$
to $(\pi,0)$ lines. The spin-flux model at mean-field exhibits another
smaller peak at $(0,\pi/2)$ which is not resolvable in existing
experimental data. This minor discrepancy may be due to  next
nearest neighbor hopping or other aspects of the electron-electron
interaction which we have not yet included in our model. \cite{11} The
quasi-particle dispersion relation of the conventional phase has a
large peak at $(\pi/2,\pi/2)$ on the $(0,0)$ to $(\pi,\pi)$ line (see
Fig. 2), but it is perfectly flat on the $(0,\pi)$ to $(\pi,0)$
line. Also, it has a large crossing from the upper to the lower
band-edge on the $(0,0)$ to $(0,\pi)$ line. This dispersion relation
is very similar to that of the $t-J$ model (see Ref. 22).
	
	The self-consistent solutions of Eqs. (\ref{2.2}),
(\ref{2.3}), (\ref{2.5}) and (\ref{2.6}) are shown in Figs. 3(a) and 3(b)
(continuous lines). Fig. 3(a) shows the magnitude of the staggered spin
as a function of $U/t$. In the large $U/t$ limit $S$ goes to 1/2, as
expected. In the small $U/t$ limit there is a solution with
$S\rightarrow 0$ only for the conventional phase. The spin-flux phase
admits an AFM ($S \neq 0$) mean-field solution solution only for $U/t
> 3$.  The ground-state energies per site are shown in Fig. 3(b), as a
function of $U/t$.  The energy of the spin-flux phase is lower than
the energy of the conventional phase, suggesting that spin-flux
provides a better mean-field starting point from which to describe
fluctuation effects on the system.

 In the large $U/t$ limit, the Hubbard model at half-filling is
equivalent with the Heisenberg model. \cite{13} This equivalence
remains true in the presence of spin-flux since the Heisenberg
exchange coupling involves only the product of phase factors
$T^{ij}T^{ji}=1$:
$$
{\cal H}={4 t^2\over U} \sum_{\langle i, j\rangle}^{} |T^{ij}|^2
\left( \vec{S}_i \vec{S}_j-{1\over 4}\right).
$$
This equivalence between the conventional and spin-flux phases is
indeed observed in all our numerical simulations when the electron
concentration is at, or extremely near, half-filling. The small
differences are due to higher order virtual hopping corrections to the
Heisenberg model. The most significant differences between the
spin-flux AFM and the conventional AFM occur at intermediate $U/t$
values.

The analytical results described above provide a useful check for our
self-consistent numerical scheme. The circles and diamonds of Figs. 3(a)
and 3(b) show the numerical results obtained in the ``bulk'' limit, in
good agreement with the analytic results. If we use Cyclic Boundary
Conditions (CBC), which require an even value for $N$, the ``bulk''
limit is reached for $N \ge 10$.

\section{Solitons in the doped Mott insulator}

When charge carriers are added to the system, the Hamiltonian depends
directly on the doping charge through the $Q(i)$ parameters (see
Eq. (\ref{1.4})).  As a result we have different mean-field
Hamiltonians for hole-doped and electron-doped systems. However, the
charge-conjugation symmetry is preserved in the sense that the
self-consistent spin and charge distributions, the one-electron
spectrum, and total energy of the hole-doped and electron-doped system
are very simply related to one another.  The correspondence is as
follows.  Let ${\cal H}^{h}$ be the Hamiltonian of a hole-doped system
defined by the parameters $\vec{S}^h(i)$ and $Q^h(i)=1-\rho(i)$, where
$\rho(i)$ is the charge distribution of the doping holes. If this
Hamiltonian is self-consistent, so is the Hamiltonian ${\cal H}^e$ of
the electron-doped system defined by the parameters
$\vec{S}^e(i)=-\vec{S}^h(i)$ and $ Q^e(i)=1+\rho(i)=2-Q^h(i)$.  This
follows from the fact that if $\phi^h(i)$ is a spinor such that ${\cal
H}^h \phi^h(i) = E \phi^h(i)$, then
$\phi^e(i)=(-1)^{(i_x+i_y)}\phi^h(i)$ satisfies ${\cal H}^e \phi^e(i)
= - E \phi^e(i)$. In other words, the doping charges are distributed
identically (only with different signs) and the final spin
configurations are identical, while the electronic spectrum of the
hole-doped system is obtained from that of the electron-doped system
by reflection with respect to $E=0$.  Also, if $n$ is the number of
charge carriers (measured with respect to the half-filled system), the
energy of the hole-doped and the electron-doped configurations are
related by $E_{hole}(n)= E_{electron}(n)-Un$.  This difference in the
energies of the equivalent hole-doped and electron-doped
configurations is entirely an artifact of the absence of a charge term
describing the interaction of the electrons with the neutralizing
positive background of nuclei.  A hole-doped configuration always
appears energetically less expensive than the corresponding
electron-doped configuration, since the latter has additional
electron-electron repulsions, with no compensating electron-nuclei
attraction.  A very simple way of compensating for this is to identify
the average energy $\left(E_{hole}(n)+ E_{electron}(n)\right)/2$ with
the energy of the state with $n$ doping charges.

	All the spin and charge distributions, as well as electronic
spectra presented in the rest of this article are the ones associated
with the corresponding hole-doped systems. For the energies of these
configurations we give the average value identified above, unless
otherwise stated. However, we emphasize that the difference between
the energies of the corresponding hole-doped and electron-doped
systems, $Un$, is independent of the distribution of the $n$
charges. Therefore, if we compare different configurations
corresponding to the same doping and $U/t$ parameter, the hole-doped,
electron-doped and averaged energies lead to the same optimum
Hartree-Fock soliton configuration.

\subsection{The spin-bag}

	If we introduce just one hole in the plane, the
self-consistent solution we get is a conventional polaron or
``spin-bag'' (see Figs. 4 and 5).  The doping hole is localized around
a particular site, leading to the appearance of a small ferromagnetic
core around that site. The spin and charge distribution at the other
sites are only slightly affected. In fact, the localization length of
the charge depends on $U/t$, and becomes very large as $US \rightarrow
0$, since in this limit the Mott-Hubbard gap closes.  For intermediate
and large $U/t$, the doping hole is almost completely localized on the
five sites of the ferromagnetic core. The static spin-configuration
surrounding the hole makes charge transport very difficult since
motion of the hole outside the ferromagnetic core will create a string
of antiferromagnetic bond defects.  The hole may circumvent this
self-trapped configuration by further twisting the AFM
background. However, the subgap electronic level induced by the
spin-bag ensures that it has a lower Hartree-Fock energy that a hole
in the valence band of a spiral (twisted) magnetic background state.

	The spin-bag is a charged fermion, as can be seen by direct
inspection of its charge and spin distributions(see Figs. 4 and
5). The electronic spectrum in the presence of the hole-doped spin-bag
(see Figs. 6(a),6(b)) reveals that two levels are drawn deep into the
Mott-Hubbard gap. These are the first empty levels, suggesting that
one of the discrete gap levels emerged from the upper edge of the
valence band, while the other one emerged from the lower edge of the
conduction band. There are also an odd number of occupied discrete
levels which split from the lower edge of the valence band (one in the
spin-flux case, three in the conventional phase).  This means that the
valence band continues to have an even (paired) number of levels, and
therefore its contribution to the total spin is zero. However, the
excitation carries the spin localized on the occupied discrete levels.
Since there is an odd number of such levels, the spin of the
excitation is a half-integer spin.

	We define the excitation energy of a spin-bag as the
difference between the energy of a self-consistent configuration with
a spin-bag and the energy of the undoped AFM background. In Fig. 7 we
show the variation of this excitation energy with the size $N$ of the
lattice, for $U/t=6$, for both cyclic and free boundary conditions. As
expected, the excitation energy of the spin-bag does not depend on the
size of the lattice, for $N\ge 10$.  The variation of the excitation
energy of the spin-bag with $U/t$ is shown in Fig.  8. In the very large
$U/t$ limit, this energy goes asymptotically from above to $U/2-2t$.
This can be understood from the fact that in this limit, a hole-doped
spin-bag should cost no Coulomb energy, since we simply remove an
electron from a site. An electron-doped spin-bag, on the other hand,
costs $U$, since we have a doubly occupied site. In both cases, the
doping charge can move within the ferromagnetic core, lowering its energy
by $2t$.  The average energy, therefore, is $U/2-2t$ as obtained
numerically.

\subsection{The meron-vortex}

In the previous discussion it was suggested that the charged spin-bag
is relatively immobile in the AFM background whereas a twisted
magnetic background would facilitate electrical conductivity.  In this
section we present another self-consistent charged soliton, the
meron-vortex (see Figs 9,10).  This excitation has a topological
(winding) number 1 (i.e. the spins on each sublattice rotate by $2\pi$
on any closed contour surrounding the center of the meron). As such,
this excitation cannot appear alone in an infinitely extended AFM
plane by the introduction of a single hole into the plane. From a
topological point of view, this is so because the AFM background has a
winding number 0, and the winding number must be conserved, unless
topological excitations migrate over the boundary into the considered
region.  Moreover, the excitation energy of the meron-vortex diverges
logarithmically with the size of the lattice. This means that an
isolated hole introduced in the AFM plane is initially dressed into a
spin-bag excitation.  Nevertheless, we study the characteristics of
the isolated meron-vortex, since this provides a foundation for
understanding multiple meron-antimeron configurations at higher
dopings, which are no longer topologically or energetically forbidden.

In order to get a self-consistent meron solution, we start with a spin
configuration with a winding number of unity. Successive iterations
conserve this winding number, but adjust the magnitude of the spins
and distribution of charges until self-consistency is reached. In this
case, it is useful to use free boundary conditions, since cyclic
boundary conditions would distort the spins near the edges of the
sample such that they orient in the same direction with the spins on
the opposite edge, affecting the excitation energy.

	From Figs. 9 and 10 we can see that the meron-vortex is a
charged boson, since the total spin of such a configuration is zero,
while it carries the doping charge. Its electronic spectrum 
is shown in Fig. 11(a),11(b). In the presence of
the hole-doped meron-vortex we see a pair of levels drawn deep into
the gap.  In the conventional AFM state these two levels are
degenerate, whereas in the spin-flux phase the degeneracy is
lifted. This is a direct consequence of the fact that the
self-consistent meron-vortex of the spin-flux phase is localized at
the center of a plaquette (as shown in Fig. 9) while a self-consistent
meron-vortex in the conventional state is localized at a site.  If the
charge dependent terms are removed from the meron-vortex Hamiltonian,
this pair of levels is exactly at the mid-gap of the Mott-Hubbard gap
for any value of $U/t$, as predicted in Reference 11.  These two
levels are the first unoccupied levels, suggesting that one of them
emerges from the valence band, while the other one emerges from the
conduction band. Moreover, they split from the $(\pi/2,\pi/2)$ peaks
of the electron dispersion relation (the Fermi points of the spin-flux
phase). \cite{3,3b} This process is consistent with the opening of
the hole pockets near $(\pi/2,\pi/2)$ in the underdoped cuprates.

The bosonic nature of the meron-vortex can be inferred from its
electronic spectrum as well. In this case (see Fig. 11(a),11(b)) only the
extended states of the valence band are occupied, and therefore they
are the only ones contributing to the total spin. Since only one state
is drawn from the valence band into the gap, becoming a discrete bound
level, it appears that an odd (unpaired) number of states was left in
the valence band. However, one must remember that for topological
reasons, merons must appear in vortex-antivortex pairs. Thus, the
valence band has an even number of (paired) levels, and the total spin
is zero. This argument of the bosonic character of the meron-vortex is
identical to that for the charged domain wall in polyacetylene. \cite
{3b,4,5}

	The excitation energy of the meron as a function of the
lattice size is shown in Fig. 12, for a fixed $U/t$. This excitation
energy was obtained by substracting the energy of an AFM undoped
background (with free boundary conditions) from the energy of the
meron-configuration.  As in the case of the spin-bag excitation, it is
energetically more expensive to excite a meron in the conventional
phase than in a spin-flux phase, for all possible values of $U/t$. The
dependence of the excitation energy on $N$ may be fitted to the
expected form $E_{meron}(N)=\alpha\ln{N} +\epsilon_{core}$. The
dependence of $\alpha$ and $\epsilon_{core}$ on $U/t$ are shown in
Figs. 13 and 14.  Both vanish as $S\rightarrow 0$ (corresponding to
$U\rightarrow 3t$ for the spin-flux phase and $U \rightarrow 0$ for
the conventional phase). In the very large $U$ limit, $\alpha
\rightarrow 0$ and $\epsilon_{core} \rightarrow U/2$ as expected. In
this limit all possible spin configurations become degenerate
(i.e. there is no difference between the excitation energy of a meron
and the excitation energy of a spin-bag). In the intermediate $U/t$
region, the core energy of the meron-vortex in the spin-flux phase is
energetically less expensive than that of the meron vortex of the
conventional phase due to the spreading of its charged core over the
four sites of a plaquette.

Comparing the energy of a meron in a finite size sample with that of a
spin-bag, we can obtain a crude estimate of the critical doping
concentration at which a transition from the spin-bags to a liquid of
charged meron-vortices may take place.  Comparing Fig. 14 with Fig. 8,
we see that $\epsilon_{core} < E_{spin-bag}$ for small and intermediate
$U/t$. This means that the excitation energy of a meron-vortex is
smaller than that of a spin-bag, provided that the effective size
$N_{eff}$ of the meron is smaller than $N_o$ defined by
$E_{meron}(N_o)= \alpha \ln(N_o)+\epsilon_{core}=E_{spin-bag}$.  The
effective size is given by the sample size $N$ or the distance to the
core of the nearest anti-meron, whichever is smaller. This suggests
that for an infinite lattice and finite doping, meron-vortex
excitations have lower energies than spin-bag excitations, provided
that each hole is dressed by a meron or antimeron-vortex and that the
average separation between the vortex and the antivortex is less than
$N_o$.  Clearly, this may occur if the doping concentration $\delta$
is larger than the critical value $\delta_c\equiv 1/N_o^2$.  We plot
this critical concentration as a function of $U/t$ in Fig. 15, for
both conventional and spin-flux phases.

	In the conventional phase we see that the purported critical
 concentration for the dissociation of spin-bags into charged
 meron-antimeron pairs is larger than 0.30. At such large doping
 concentrations the average size of the excitation is $N_o < 2$, and
 the distinction between merons and spin-bags is blurred.  We
 conclude, therefore, that there is no clear transition from a state
 with spin-bag excitations to a state with meron excitations as the
 doping increases. In other words, the only relevant excitations for
 the conventional AFM phase are spin-bags, within the Hartree-Fock
 approximation.

	In the spin-flux phase, the situation is very different. For a
broad range of intermediate values of $U/t$ the critical concentration
$\delta_c$ is small and the distinction between spin-bags and merons
remains clear. This suggests that for these values of $U/t$ there are
two distinct types of Hartree-Fock ground-states, as a function of
doping. At very low dopings, the spin-bag excitations are
energetically more favorable. Since spin-bags affect the magnetic
order only locally, the long range AFM order is still preserved in
their presence.  However, if the concentration increases beyond
$\delta_c$, it is energetically favorable for each hole to be
surrounded by a meron or antimeron-vortex. In this case, the long
range AFM order can be destroyed, leaving behind either power-law
decaying magnetic correlations or short range AFM on the length scale
of the average distance between vortices.

In the above estimate of the critical concentration $\delta_c$ we
 assumed that the merons and antimerons are uniformly
 distributed. However, the actual critical concentration $\delta_c$
 may be lowered when the tendency of merons and antimerons to form
 tightly bound pairs (of total winding number 0) is considered. In
 Figs. 16 and 17 we show the self-consistent spin and charge
 distributions for the lowest energy configuration found when we put 2
 holes on the AFM lattice. It consists of a meron and an antimeron
 centered on neighboring plaquettes. As a result of the interaction,
 the cores of the vortices are somewhat distorted, and most of the
 charge missing from the (10,10) site which is common to both
 cores. If the vortices were uncharged, a total collapse of the
 vortex-antivortex pair would be plausible.  However, for charged
 vortices, the fermionic nature of the underlying electrons prevents
 two holes from being localized at the same site, in spite of the
 bosonic character of the collective excitation.

A very interesting feature of this tightly bound meron-antimeron solution
is that the attraction between the charged vortices is of purely 
topological nature, and appears even though the electronic Hamiltonian (1)
contains only repulsive electron interactions. Vortex-antivortex
attraction varies as the logarithm of the distance between the cores,
and therefore the pair of vortices should remain bound even if full
Coulomb repulsion exists between the charged cores. Thus, the process
of nucleation of meron-antimeron pairs upon doping provides a very
natural scenario for the existence of pre-formed pairs in the
underdoped regime.

There is another possible self-consistent state for the system with
two holes,
consisting of two spin-bags far from each other (such that their
localized wave functions do not overlap). The excitation energy of
such a pair of spin-bags is simply twice the excitation energy of a
single spin-bag.  When this excitation energy is compared to the
excitation energy of the tightly bound meron-antimeron pair, we find
that it is higher by $0.15t$ (for $U/t=5$). This would suggest that
spin-bags are always unstable to the creation of charged
meron-antimeron pairs within the spin-flux phase, and that the
critical doping concentration, $\delta_c$, should be set equal to
zero. A more realistic determination of the critical hole
concentration for the nucleation of meron-antimeron pairs requires the
incorporation of the long range Coulomb repulsion between charge
carriers in the doped Mott insulator.

	The situation in the high-T$_c$ copper-oxide materials is
probably more complex, and depends on the nature of the doping
process. If the charge carrier concentration is low and uniformly
distributed, the average distance between holes is large. At low
temperatures, it is possible that these holes are trapped somewhere in
the vicinity of their donors in the form of spin-bags. If two
spin-bags encounter each other, they should indeed decay into a
tightly bound meron-antimeron pair.  Since such a pair distorts the
AFM background only in a very small region, magnetic LRO is preserved.
At low temperatures and low dopings, these meron-antimeron pairs may
remain pinned to the donor atoms or other forms of disorder, giving
rise to the appearance of a spin-glass type phase of the magnetic
background.  At higher dopings the pinning potential of the donor
atoms is screened and the soliton-soliton interactions are stronger
than pinning energies.  For concentrations greater than some critical
concentration, it is possible that charged meron-antimeron pairs are
no longer tightly bound and AFM long range order is completely
destroyed.

	If this scenario is applicable to the high-T$_c$ copper oxide
materials, it is tempting to associate the charge carriers in the
doping regime relevant to superconductivity with meron-vortices.
Besides the magnetic order, another extremely important issue is the
dynamics of solitons. For instance, in the intermediate $U/t$ regime,
a spin-bag as depicted in this model (see Fig. 4,5) is basically
immobile, since moving would mean leaving behind a string of
ferromagnetically aligned spins. It is plausible \cite{14} that
the kinetic energy of localization of the hole could be lowered if the
spin-bag (spin-polaron) has a ferromagnetically aligned core, within
which the hole is free to move. Another possibility \cite{15} is that
a spiral twist in the AFM background accompanies the hole as it moves.
The meron-vortex may be regarded as a self-consistent realization of
the twist-accompanied hole which is topologically stable even when the
charge carrier is stationary.  The vortex in the AFM background
surrounding the hole facilitates mobility of charge since hopping of
the vortex core to a neighboring plaquette leads to a less severe
distortion of the AFM exchange coupling between neighboring spins.
Since meron-vortices have a bosonic nature, the non-Fermi liquid
nature of the  metal from which superconductivity emerges is
also quite natural.

\section{Higher dopings: multi-soliton configurations}

For higher carrier concentrations, there is some arbitrariness in
choosing the initial spin and charge configurations from which to
begin the iterative self-consistency scheme.  Since a variety of
different self-consistent states may be realized starting from
different initial configurations, we adopt a probabilistic
approach. We give random numbers as the initial components of the spin
distribution, and also choose randomly the sites where the holes are
initially localized. This iterative process is repeated many times,
and the self-consistent configuration of lowest energy is finally
selected.  As we mentioned before, the relation between the energy per
site of the hole-doped configuration and that of the equivalent
electron-doped configuration is given by
$e_{hole}(\delta)=e_{electron}(\delta)-U\delta$ where $\delta$ is the
average number of charge carriers per site.  For convenience, we plot
the energy of the hole-doped configurations as a function of
doping. The results obtained in the random searches are summarized in
Fig. 18, where the energy per site (in units of $t$) of the
self-consistent hole-doped configurations is plotted as a function of
the electron concentration. They correspond to $U/t=5$, and a $10
\times 10$ lattice with cyclic boundary conditions.  For $\delta=0.01$
and $ 0.02$ (corresponding to one and two holes, respectively) we
recapture the results presented in the previous section. In the
spin-flux phase, a single isolated hole forms a spin-bag, whereas the
lowest energy configuration found for two holes is the tightly bound
meron-antimeron pair shown in Figs. 16,17.  For 2 holes we also find a
number of self-consistent metastable states, containing widely
separated spin-bags.

For a doping up to about $0.30$, (which corresponds to an electron
density $c=0.70$ to $1$), the lowest energy configurations always
correspond to various arrangements of meron-antimeron pairs in the
spin-flux phase.  As an example, we show the spin configuration of lowest
energy found for 8 holes, $c=0.08$ (see Fig. 19). We can see 4
meron-antimeron pairs arranged such that each meron is surrounded by
antimerons and viceversa. This state appears to be a crystal of
meron-antimeron pairs, in the sense that the lattice obtained through
translations of the 10x10 lattice shown in Fig. 19 has an ordered
distribution of meron-antimeron pairs.  This crystallization is very
likely an artifact of the zero temperature, semiclassical, static
Hartree-Fock approach. Incorporation of dynamics and fluctuations in
the model may lead to the melting of this crystal into a quantum
liquid of mobile meron-antimeron pairs. However, as discussed in the
next section, ordered arrays of merons and antimerons may play an
important role at some special dopings.

As the doping becomes larger than $0.20$, the density of vortices is
so large that we obtain configurations which have a meron or antimeron
vortex localized on almost each plaquette, leading to a state where
long range AFM correlations are completely lost. As
a consequence, the doping charge is quite uniformly distributed over
the whole lattice, and the magnitude of the staggered spin decreases
considerably. The apparent overlap of the charge carrier
wavefunctions, here, suggests that quantum corrections to the
Hartree-Fock approximation may be substantial in this doping range and
that the charge carriers here form a novel type of quantum liquid.
Finally, at very large doping ($\delta > 0.4$), the entire spin-flux
phase is energetically unstable to the formation of a conventional
electron gas.  This is expected, since (see Eqs. (\ref{2.1}) and
(\ref{2.4})) the bottom of the valence band in the spin-flux phase is
higher than that of the valence band in the conventional phase. At
very low electron concentrations, the energy per site approaches that
of the non-interacting ( $U=0$) electron model (see Fig. 18), as
expected.

	In contrast to the above picture of a meron-liquid in the
spin-flux phase, the lowest energy configurations at low dopings in
the conventional AFM phase always consists of charged
stripes. \cite{16,16bb} For example, in Fig. 20 we show the
selfconsistent spin 
configuration found for $c=0.15$, where all the spin-bags assemble in
a closed stripe (the stripe must close due to the cyclic boundary
conditions). We have also calculated the energies of ordered
horizontal and diagonal stripes. This necessitates bigger lattices, so
that the cyclic boundary conditions are satisfied. For $U/t=5$ we find
basically no difference between the energies of such stripes and that
of the closed stripes obtained from the random initial conditions.
The instability of the spin-bags to stripe formation has also been
proven in the three-band Hubbard model and the t-J model. \cite{16bis}
However, as seen from Fig. 18, such states have higher energies than
the liquid of meron-antimeron pairs of the spin-flux phase.  We have
also tried to obtain horizontal and diagonal stripes in the spin-flux
phase, by starting with an initial configuration containing such a
stripe. However, they converge to the liquid of meron-antimeron pairs
rather than self-consistent stripe configurations.

At large dopings, the conventional phase becomes energetically more
favorable and more and more discrete levels are drawn into the
Mott-Hubbard gap. Due to overlap of the charge carrier wavefunctions,
these levels spread into a broad impurity-like band. Also, the
staggered magnetization $S$ at each site is strongly suppressed,
leading to shifts of the band edge energies (roughly given by $\pm
US$). These two effects conspire to close the Mott-Hubbard gap, and
lead to the formation of a conventional, Fermi-liquid with a
partially-filled band for dopings $\delta > 0.3$.

In summary, our picture is that of three distinct regimes. At very low
dopings, we have a collection of tightly bound meron-antimeron pairs
and/or spin-bags, which preserve the long range AFM order. When the
doping exceeds some critical value $\delta_c$, a transition to a
quantum liquid of meron-antimeron pairs occurs, and is accompanied by
the destruction of the AFM long range order. Since these charged
merons are spinless bosons this metallic state will invariably exhibit
non-Fermi-liquid properties. As the doping further increases, the
spin-flux state itself is unstable, the Mott-Hubbard gap closes and
the system reverts to a conventional Fermi-liquid. Although our static
Hartree-Fock analysis points to the above picture, it does not
describe soliton dynamics and quantum fluctuation effects pertinent to
the quantum liquid phase of merons. In addition, a more careful
treatment of the long range part of the Coulomb repulsion between
charge carriers may be needed in the non-Fermi-liquid phase where the
conventional arguments of screening are inapplicable. \cite{17}
Finally, a more realistic model must include the interactions of the
doping charges with the impurity charges located in nearby planes, and
the influence of structural distortions of the CuO planes. The last
issue is pertinent to the meron crystal phase at the special doping
$\delta=1/8$, which we discuss below.

\subsection{Charge carrier concentration of $\delta=1/8$}

	The $\delta=1/8$ doping is very special, because in some
compounds \cite{18} superconductivity is completely suppressed at this
doping. A very simple and natural explanation of this suppression is
that the charge carriers become immobile. Within our picture of a
charged meron liquid, at a doping of 1/8, we find a self-consistent
structure consisting of a crystal of merons and antimerons (see
Figs. 21,22). Neutron scattering reveals that for $\delta=1/8$, the
${\pi\over a}(1,1)$ AFM magnetic peak splits into four peaks situated
at ${\pi\over a}\left( 1 , 1\pm {1\over 8}\right)$ and ${\pi\over
a}\left( 1 \pm {1\over 8}, 1\right)$. \cite{19,20} For the calculated
meron crystal shown in Figs. 21,22, the magnetic structure factor
exhibits four peaks with the correct distances between the
peaks. However, they are rotated by $45^{\circ}$ (they appear along
the diagonals, not along the horizontals) relative to the observed
neutron scattering peaks. This picture can be brought into agreement
with experiments by introducing a small anisotropy in the electron
hopping within the copper-oxide plane. The addition of such a
perturbation to our model is justified by the experimentally observed
distortion of the lattice from the usual low-temperature orthorhombic
(LTO) structure to the low-temperature tetragonal (LTT) structure at
this doping.  \cite{19bis} In the LTT structure the atomic
displacements form a horizontal (or vertical) structure, and very
likely favor the pinning of horizontal (vertical) stripes. The easiest
way to mimic this structural distortion is to add a small anisotropy
in the magnitude of the hopping integral, with the same
periodicity. For a $3\% $ anisotropy, the half-filled stripe structure
predicted by Tranquada \cite{20} becomes stable (see Fig. 23).
The self-consistent stripe configuration obtained in the presence of
the small anisotropy is made up of merons and antimerons packed along
horizontal lines.  This example illustrates that a more realistic
model including the effects of such structural distortions and
possibly the interaction with the impurity charges is required for a
quantitative comparison with experiments.

\subsection{Optical absorption}

As we mentioned before, in the presence of each meron or
antimeron-vortex, two electronic levels, one from the valence band and
one from the conduction band, are drawn deep into the gap.  In the
presence of multi-soliton interactions these levels spread into a
broad impurity band within the larger Mott-Hubbard gap.  Since these
localized states are empty (for the hole-doped system), electrons can
be optically excited to them from the valence band. Consequently, a
broad optical absorption band appears inside the Mott-Hubbard gap. In
Fig. 24 we show the evolution of the optical absorption with
doping. The absorption was calculated through straightforward
perturbation theory, after a term coupling the doping charge to an
external vector field \cite{new} was added to the mean-field
Hamiltonian. This leads to the following formula for the electric
conductivity tensor
$$
\sigma_{ij}(\omega)={1\over i\omega} \sum_{\alpha=1}^{N_e}
\sum_{\beta=N_e+1}^{N^2} [ {\eta^i_{\alpha\beta}
\eta^j_{\beta\alpha} \over \hbar\omega-(E_{\alpha}-E_{\beta})+i\Gamma}
$$
$$
-  {\eta^i_{\beta\alpha}
\eta^j_{\alpha\beta} \over \hbar\omega+(E_{\alpha}-E_{\beta})+i\Gamma}
]
$$
where 
$$\vec{\eta}_{\alpha\beta}={i e t\over \hbar} \sum_{\langle i, j
\rangle }  
[ (\vec{r}_j-\vec{r}_i)\phi_{\alpha}^*(i) T^{ij} \phi_{\beta}(j)
$$
$$
+ 
(\vec{r}_i-\vec{r}_j)\phi_{\alpha}^*(j) T^{ji} \phi_{\beta}(i) ]
$$
is the matrix element between an occupied state $\alpha$ and an empty
state $\beta$ of the density of current operator $\vec{j}$. Here,
$N_e$ is the number of occupied states, $N^2$ is the total number of
states and $\Gamma$ is a phenomenological damping coefficient.  The
calculation is approximate, in the sense that we did not include the
variation of the spin and charge distribution due to the modification
of the wave-function in the external field. A more detailed
calculation involving a time-dependent generalization of the
Hartree-Fock method (the random phase approximation) will be presented
elsewhere.  In order to mimic the interaction with spin-waves and
other damping effects on the excited electronic states we assumed that
the subgap levels are homogeneously broadened with a spectral width of
$\Gamma=0.1 t$. This leads to a smooth optical absorption even for a
$10\times10$ lattice.  As the doping increases, two effects are
apparent. The first is the appearance of a broad absorption band deep
into the gap. This is due to the soliton gap states, and significant
weight is transferred into it from the conduction band.  The second
effect is that the overall charge-transfer gap itself decreases since
doping decreases the average self-consistent value of $S$, leading to
a shift in the mean-field position of the valence and conduction band
edges. This second effect is less apparent in Fig. 24, because of the
fairly large damping we chose.  It is well known that the optical
absorption of the doped compounds contains a broad mid-infrared band
and a Drude-like tail starting from very low frequencies. \cite{f} It
is reasonable to associate the broad mid-infrared band with electronic
transitions from the valence band to the vortex mediated impurity band
(see Fig. 24). On the other hand, the Drude tail component is
associated with the translational motion of the bosonic
meron-vortices. This may be described in a time-dependent Hartree-Fock
(TDHF) approximation.

\subsection{The magnetic structure factor}

We can characterize the evolution of the magnetic long range order
with the doping by looking at the magnetic structure factor. We have
calculated the static magnetic structure factor,
$$
S(\vec{Q})={1\over N^2}\sum_{i,j}^{}e^{i\vec{Q}\cdot
\left(\vec{r}_i-\vec{r}_j\right)}\vec{S}(i)\vec{S}(j)	,
$$
assuming that the spins are frozen in the self-consistent
configuration.  The results are shown in Fig. 25. The AFM parent
compound has a large peak at $(\pi/2,\pi/2)$, as expected. As the
doping increases, this peak splits into four incommensurate peaks,
whose positions shift with the doping. This is in agreement with the
observed behavior of some cuprate compounds . \cite{19,20}

\section{Discussion and conclusions}

In this article we presented a numerical study of a mean-field
approximation of the one-band extended Hubbard model. We have shown
that at low dopings, the spin-flux phase provides a better starting
point than the conventional phase. For $U/t$ in the intermediate
range, the lowest energy configurations found in the doping regime
relevant to superconductivity, consists of a liquid of meron and
antimeron-vortices. These meron-vortices are mobile, charge carrying
bosons which accommodate each of the doping holes in an impurity band
that occurs within the Mott-Hubbard charge transfer gap. The key
ingredient that distinguishes our model from previous studies of the
Hubbard model is the concept of spin-flux. In its absence, our
analysis reproduces the conventional AFM, in which there is a tendency
for stripe formation at low doping, as predicted by many other
authors. \cite{16,16bb,16bis} However, introduction of spin-flux into the
AFM leads to a lower mean-field energy state, in which the doping
holes find it energetically favorable to be cloaked by vortices of the
magnetic background. This cloaking stabilizes the magnetic vortex, and
also facilitates the mobility of holes in the AFM background.  At
extremely low doping the holes are either paired in tightly bound
meron-antimeron pairs, or become spin-bags (which may be thought of as
a collapsed charged meron-neutral antimeron pair). Increasing doping
creates a liquid of meron-antimeron pairs, completely destroying AFM
order. This picture is consistent with angle-resolved photo-emission
studies of the quasi-particle dispersion relation, the appearance of a
broad mid-infrared optical absorption band with doping and various
aspects of the neutron scattering data. It also offers a microscopic
mechanism for the non-Fermi liquid characteristic of the metallic
state from which superconductivity emerges. 

One of the great challenges in the understanding of charge carrier
pairing is that an attractive force must emerge from a purely
repulsive many-electron Hamiltonian. This problem is exacerbated by
the fact that the standard arguments of screening of the Coulomb
repulsion are based on Fermi-liquid theory and may be inapplicable to
a doped Mott insulator. Our model, based on charged vortex solitons,
provides a very natural strong attractive force between charge
carriers which is of topological origin and which can lead to binding
of charge carriers even in the absence of screening. Moreover, the
presence of vortices in the AFM background will lead to a large
renormalization of the spin-wave spectrum. This may in turn be related
to the observed pseudo-gap phenomenon in the high $T_c$ cuprates. \cite{22}

 As the doping increases
further, the spin-flux phase is unstable to the formation of a
conventional Fermi liquid, in which the Mott-Hubbard gap is closed.

All of our results, so far, are based on a static Hartree-Fock
mean-field theory. In spite of the simplicity of our approximation,
the calculated properties of our model are consistent with a variety
of independent experimental signatures of the cuprate
superconductors. We believe that it is reasonable to proceed beyond
this mean-field theory to understand in greater detail the quantum and
dynamical properties of the meron-liquid.  The long range part of the
Coulomb repulsion between doping charges may play a more important
role in the properties of this novel quantum liquid than it does in a
conventional Fermi liquid where standard screening arguments
apply. Also, additional interactions such as crystal field effects and
conventional spin-orbit interaction,\cite{7bis} which help to stabilize uncharged
meron-vortices, may need to be added in the starting Hamiltonian (1).
These considerations may, in turn, shed light on the microscopic
mechanism of high-temperature superconductivity and the detailed
characteristics of the non-Fermi-liquid state from which it arises.

\section*{Acknowledgments}

We are grateful to Dr. Vladimir Stephanovich for a number of
stimulating discussions and to Dr.  Dirk Morr for drawing our
attention to the recent ARPES data for Sr$_2$CuO$_2$Cl$_2$.
M.B. acknowledges support from the Ontario Graduate Scholarship
Program. This work was supported in part by the Natural Sciences and
Engineering Research Council of Canada.

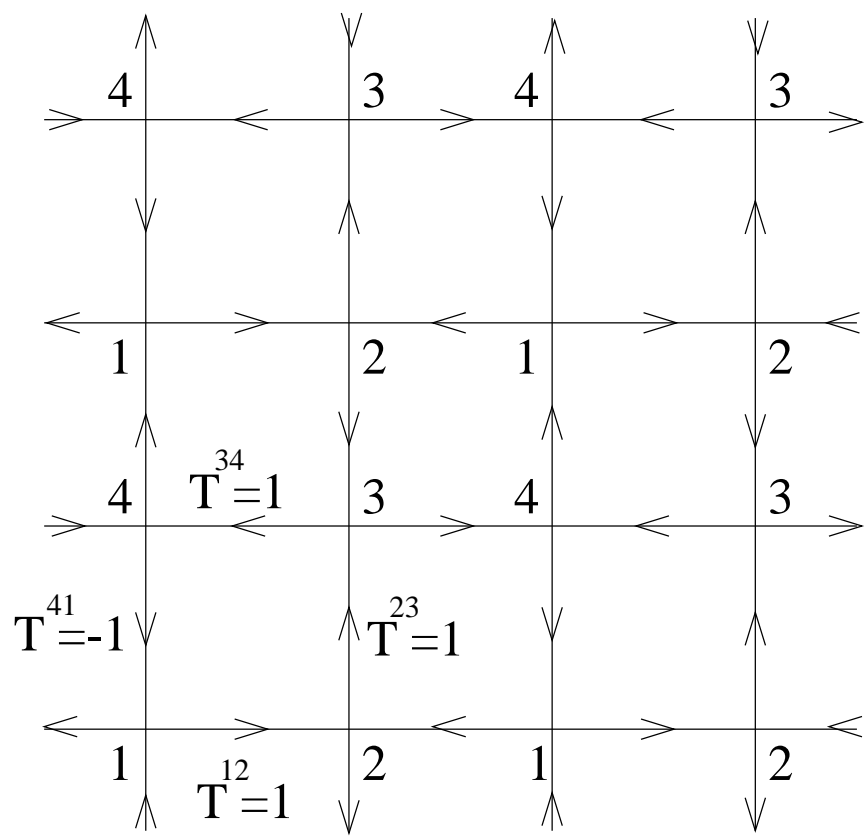
\begin{figure}
\caption{Choice of the gauge for describing the mean-field spin-flux
background. Physical observables depend on the rotation matrices
$T^{ij}$ only through the plaquette matrix product
$T^{12}T^{23}T^{34}T^{41}$. Shown above is the simplest (spin
independent) gauge choice describing a $2\pi$-rotation of the internal
coordinate system of the electron (described by 3 Euler angles) as it
encircles an elementary plaquette. This is a new form of spontaneous
symmetry breaking for a strongly interacting electron system in which
the mean-field Hamiltonian acquires a term with the symmetry of a
spin-orbit interaction.  }
\end{figure}

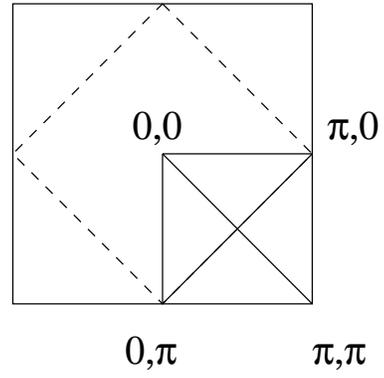
\begin{figure}
\caption{A comparison between the experimentally determined $E(\vec{k})$
quasi-particle dispersion relation, from angle resolved photoemission
studies (ARPES), for the insulating Sr$_2$CuO$_2$Cl$_2$ (open circles
with error bars) and the mean-field AFM spin-flux phase dispersion
relation (full line) and the mean-field AFM conventional phase
dispersion relation (dashed-dotted line). While the peak on the
$(0,0)$ to $(\pi,\pi)$ is equally well described in both models, the
mean-field spin-flux model gives a much better agreement for the
$(\pi,0)$ to $(0,0)$ and $(\pi,0)$ to $(0,\pi)$ directions.  The
fitting corresponds to $U=2.01$ eV, $t=0.29$ eV for the spin-flux
phase, and $U=1.98$ eV, $t=0.21$ eV in the conventional phase. The
experimental results are the ARPES results of Ref. 21.  }
\end{figure}

\begin{figure}
\caption{ (a) Dependence of the staggered spin $S$ of the AFM
undoped compound with $U/t$. The diamonds show the numerical results
obtained in the presence of the spin-flux, while circles show
numerical results for the conventional phase. The lines show the
values predicted by Eqs. (6) and (9) for the two phases. The
spin-flux phase has a mean-field AFM solution ($S \neq 0$) only for
$U/t > 3$. As
expected, $S \rightarrow 1/2$ in the large $U/t$ limit, where every
electron becomes localized on individual sites.
(b)  Dependence of the ground-state energy per site
of the AFM parent compound with $U/t$. 
The diamonds show the numerical results
obtained in the presence of the spin-flux, while circles show
numerical results for the conventional phase. The lines show the
values predicted by Eqs. (5) and (8). for the two phases. 
The mean-field AFM spin-flux phase has a lower energy than the
mean-field AFM conventional phase for all values of $U/t$.
}
\end{figure}

\begin{figure}
\caption{Self-consistent spin distribution of a 10x10 lattice with a
spin bag centered at (5,5). The spin-bag has a small ferromagnetic
core, and the magnetic order is only locally affected.  }
\end{figure}

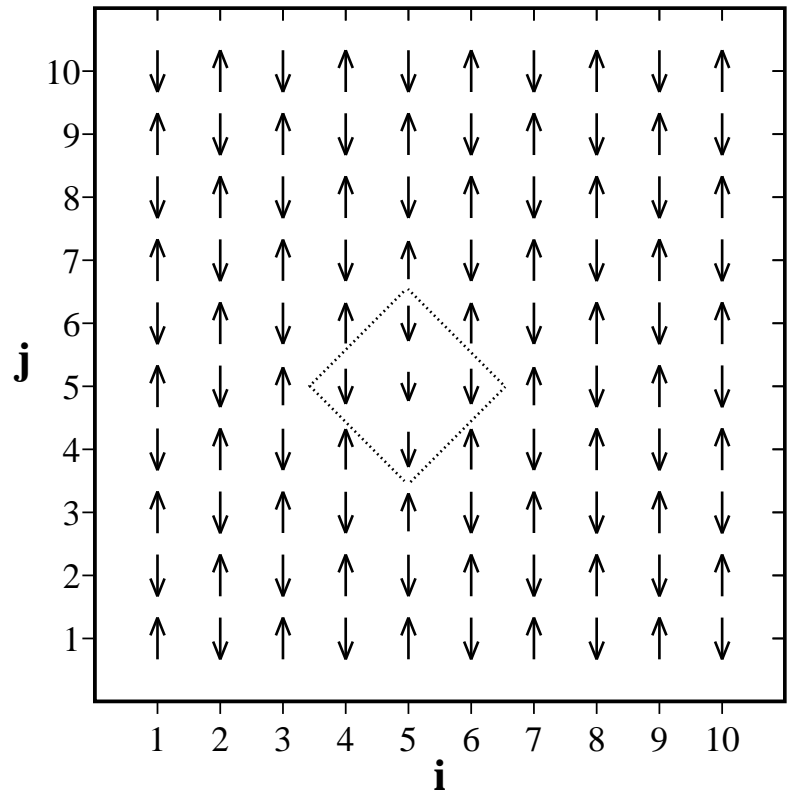
\begin{figure}
\caption{Self-consistent charge distribution of a 10x10 lattice with a
spin bag centered at (5,5). There is an average of one electron per
site everywhere, except in the core of the spin-bag where the doping
hole is localized.  }
\end{figure}

\begin{figure}
\caption{ (a) Electronic spectrum of a spin bag on a 10x10 lattice,
for $U/t=5$ and spin-flux.  Eigenenergies $E_{\alpha}$ are plotted as
a function of $\alpha=1,200(=2N^2)$. Only the first $N^2-1=99$ states
are occupied. There are two empty bound discrete levels deep into the
Mott-Hubbard gap ($\alpha=100, 101$), one of which comes from the
valence band of the undoped AFM compound (see inset). There is also an
occupied discrete level below the valence band ($\alpha=1$). The
valence band is spin paired, since it has an even number of
levels. Thus, the total spin of the spin-bag comes from the discrete
occupied level. The spin-bag is a charged, spin-${1\over 2}$ fermion.
(b) Electronic spectrum of a spin bag on a 10x10 lattice, for $U/t=5$
in the conventional state.  Eigenenergies $E_{\alpha}$ are plotted as
a function of $\alpha=1,200(=2N^2)$. Only the first $N^2-1=99$ states
are occupied. There are two empty bound discrete levels deep into the
Mott-Hubbard gap ($\alpha=99, 100$), one of which comes from the
valence band of the undoped AFM compound (see inset). There are also
three occupied discrete levels below the valence band
($\alpha=1,2,3$). The valence band is spin paired, since it has an
even number of levels. Thus, the total spin of the spin-bag comes from
the discrete occupied levels. The spin-bag is a charged, spin-${1\over
2}$ fermion.  }
\end{figure}

\begin{figure}
\caption{Excitation energy of a spin bag $E_{spin-bag}$ as a function
of the lattice size $N$. Diamonds show results for a spin-flux AFM
phase, with CBC (full diamonds) and FBC (empty diamonds). Circles show
results for the conventional AFM state, with CBC (full circles) and
FBC (empty circles). The Hubbard parameter is $U/t=6$. The bulk limit
is reached for $N >10$. In this limit, the excitation energy of the
localized spin-bag becomes independent on the size of the lattice.  }
\end{figure}

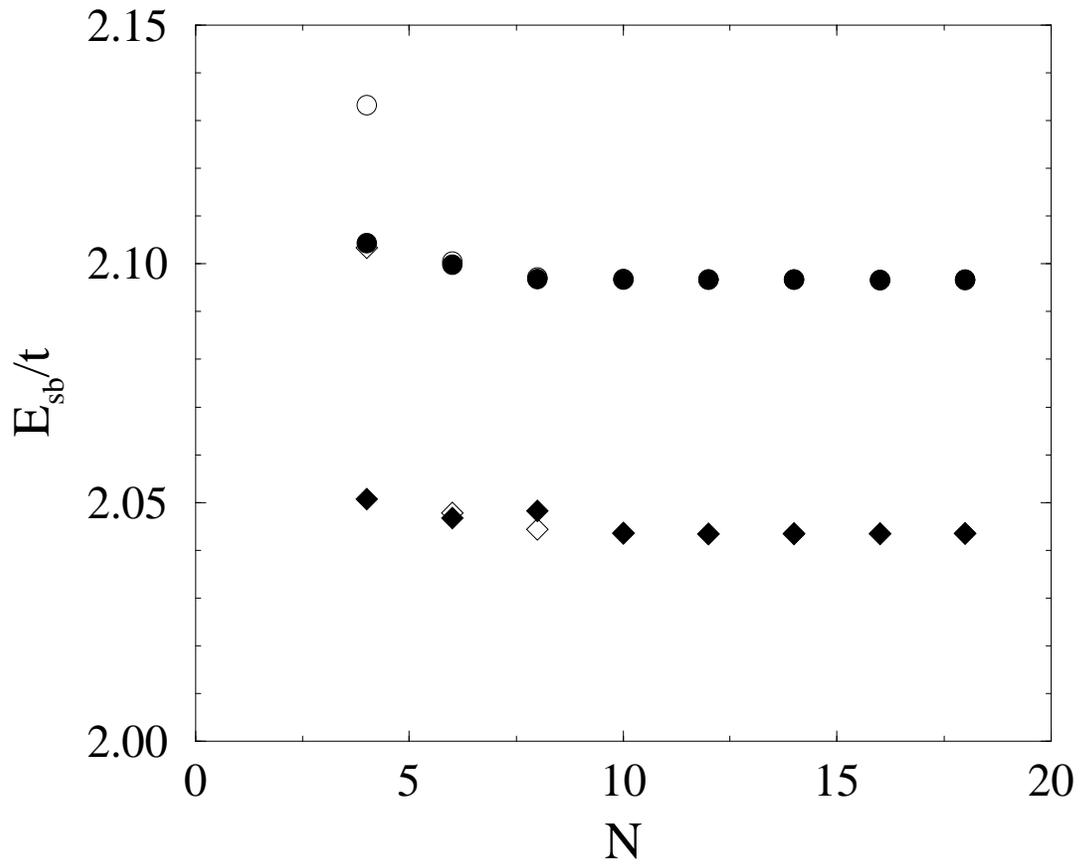
\begin{figure}
\caption{ Excitation energy of a spin bag $E_{spin-bag}$ as a function
of $U/t$, in the presence of spin-flux (filled diamonds), and in the
conventional state (circles). In the very large $U/t$ limit, the
excitation energy approaches asymptotically the value $U/2-2t$. The
excitation energy of a spin-bag is lower in the spin-flux phase than
in the conventional phase.  }
\end{figure}

\begin{figure}
\caption{ Self-consistent spin distribution of a 10x10 lattice with a
meron-vortex in the spin-flux phase. The core of the meron is
localized in the center of a plaquette, in the spin-flux phase (in the
conventional phase, the core of the meron-vortex is localized at a
site). This excitation has a topological winding number 1, since the
spins on either sublattice rotate by $2\pi$ on any curve surrounding
the core. The magnitude of the staggered magnetic moments is slightly
diminished near the vortex core but is equal to that
of the undoped AFM background far from the core.  }
\end{figure}

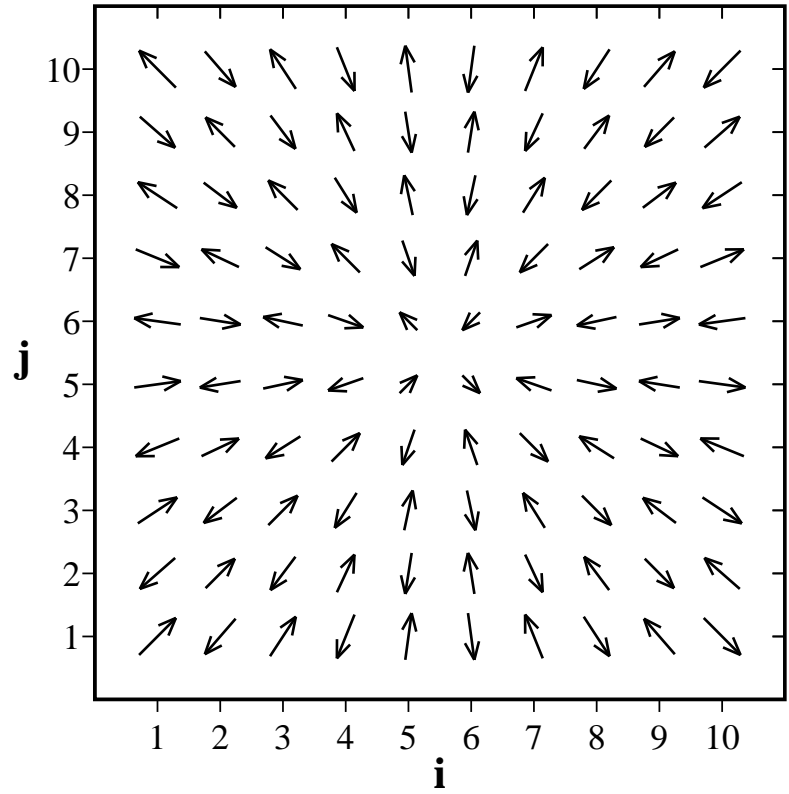
\begin{figure}
\caption{Self-consistent charge distribution of a 10x10 lattice with a
meron-vortex in the spin-flux phase. Most of the doping charge is
localized in the center of the meron. Far from the core, there is an
average of one electron per site.  }
\end{figure}

\begin{figure}
\caption{(a) Electronic spectrum of a meron-vortex on a 10x10 lattice,
for $U/t=5$, in the presence of the spin-flux. Eigenenergies
$E_{\alpha}$ are plotted as a function of $\alpha=1,200(=2N^2)$. Only
the first $N^2-1=99$ states are occupied (the valence band). There are
two discrete empty levels deep into the Mott-Hubbard gap, one of which
($\alpha=100$) comes from the valence band of the undoped AFM
parent. Merons must be created in vortex-antivortex pairs (for
topological reasons). Each pair removes two levels from the undoped
AFM valence band. Thus, the valence band remains spin paired, and the
total spin of this excitation is zero. This meron is a spinless,
charged, bosonic collective excitation of the doped antiferromagnet.
(b) Electronic spectrum of a meron-vortex on a 10x10 lattice, for
$U/t=5$ , in the conventional phase.  Eigenenergies $E_{\alpha}$ are
plotted as a function of $\alpha=1,200(=2N^2)$. Only the first
$N^2-1=99$ states are occupied (the valence band). There is a double
degenerate unoccupied bound discrete level deep into the Mott-Hubbard
gap. One of these bound levels ($\alpha=100$) comes from the valence
band of the undoped AFM parent. Merons must be created in
vortex-antivortex pairs (for topological reasons). Each pair removes
two levels from the undoped AFM valence band. Thus, the valence band
remains spin paired, and the total spin of this excitation is zero.  }
\end{figure}

\begin{figure}
\caption{Excitation energy (in units of $t$) of a single meron-vortex,
as a function of the meron size $N$, in the presence of the spin flux
(diamonds) and without spin-flux (circles). The lines show fits with a
logarithmic dependence on $N$, $E_{meron}(N)=\alpha\ln{N}
+\epsilon_{core}$. The excitation energy of a meron-vortex is always
lower in the spin-flux phase than in the conventional phase. If the
size of the meron core is small enough, the excitation energy of the
meron-vortex may become smaller than the excitation energy of a
spin-bag.  }
\end{figure}

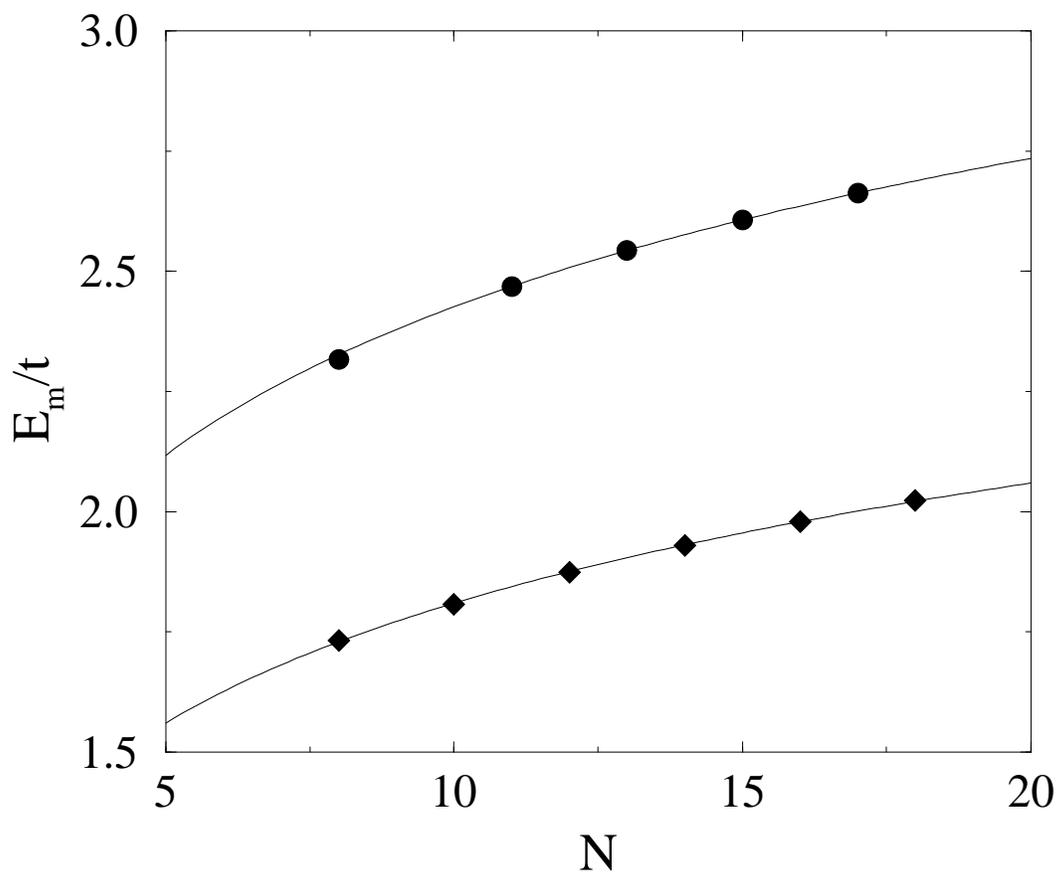
\begin{figure}
\caption{Dependence of the coefficient $\alpha$ (in units of $t$) from
the fit $E_{meron}(N)=\alpha\ln{N} +\epsilon_{core}$ on
$U/t$. Diamonds show results for a spin-flux phase, while circles
correspond to a conventional state.  The line serves to guide the
eye. In the large $U/t$ limit $\alpha \rightarrow 0$, since in this
limit all spin configurations become degenerate and the excitation
energy of the meron-vortex should equal the excitation energy of the
spin-bag.  }
\end{figure}

\begin{figure}
\caption{Dependence of $\epsilon_{core}$ (in units of $t$) from the
fit $E_{meron}(N)=\alpha\ln{N} +\epsilon_{core}$ on $U/t$. Diamonds
show results for a spin-flux phase, while circles correspond to a
conventional state.  The line serves to guide the eye. In the large
$U/t$ limit $\epsilon_{core} \rightarrow E_{spin-bag}$, since in this
limit all spin configurations become degenerate and the excitation
energy of the meron-vortex should equal the excitation energy of the
spin-bag.  }
\end{figure}

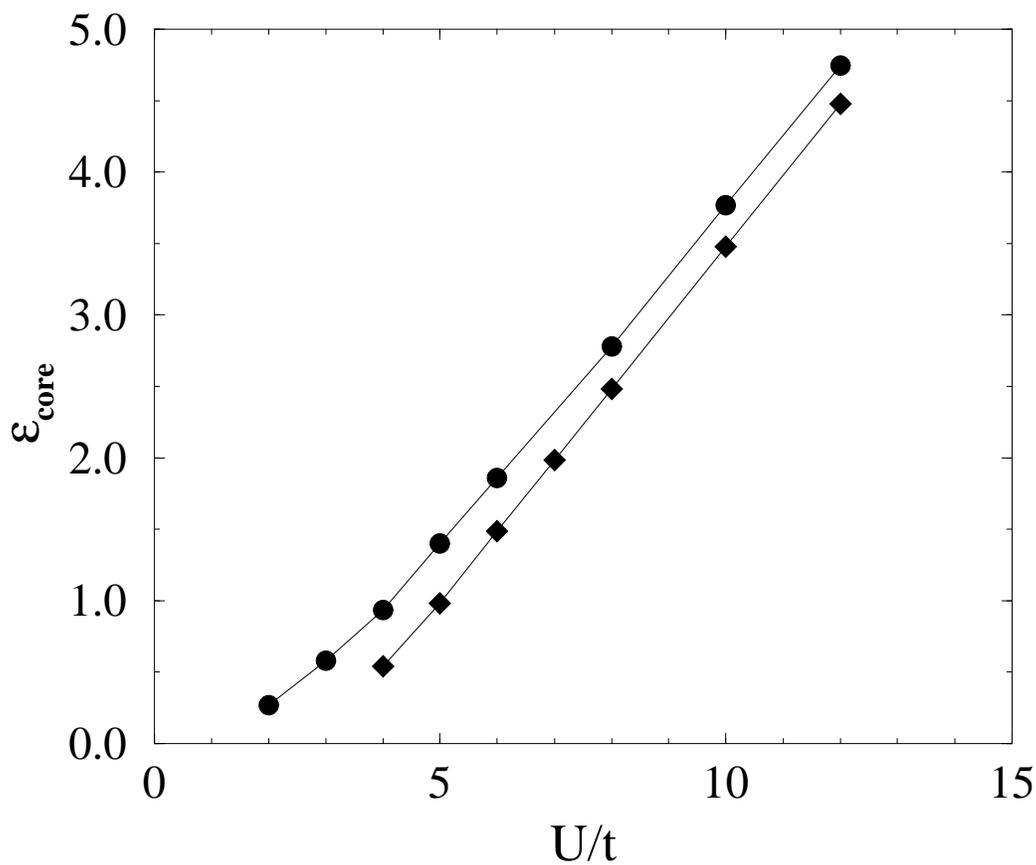
\begin{figure}
\caption{The critical doping concentration, $\delta_c$, above which a
charged meron-antimeron liquid is energetically favorable compared to
a gas of spin-bags. Diamonds show results for a spin-flux phase, while
circles correspond to a conventional state.  The line serves to guide
the eye. In the conventional phase the critical concentrations are
very large, $\delta_c > 0.3$. In the spin-flux phase, the transition
to the liquid of meron vortices takes place at dopings smaller than
$0.10$, for $U/t <8$.  In the conventional phase, the critical doping
is so large ($\delta_c > 0.30$) that merons are unlikely to appear
before the Mott-Hubbard gap itself closes.  }
\end{figure}

\begin{figure}
\caption{Self-consistent spin distribution for a tightly bound
meron-antimeron pair. The meron (M)  and the antimeron (A) are localized on
neighboring plaquettes. The total winding number of the pair is zero. The
magnetic AFM order is disturbed only on the small region where the
vortices are localized. 
The attraction between holes is of topological
nature and on long length scale is stronger than unscreened Coulomb
repulsion between charges.  }
\end{figure}

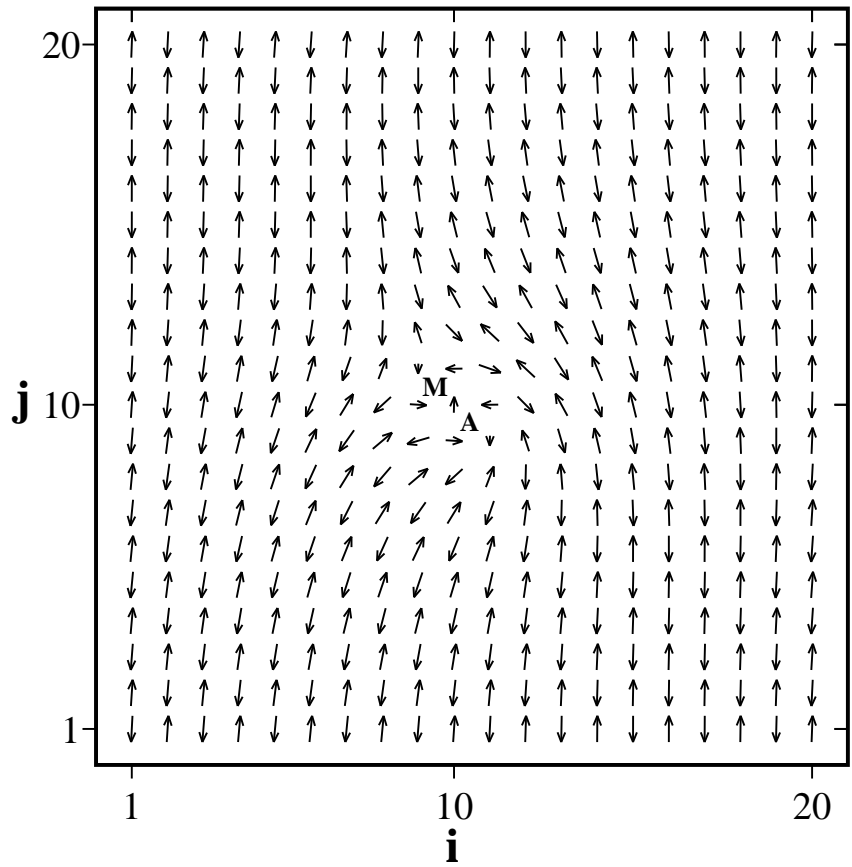
\begin{figure}
\caption{Self-consistent charge distribution for a tightly bound
meron-antimeron pair. The doping charge is mostly localized on the two
plaquettes containing the meron and antimeron cores. Due to
interactions, the cores of the vortices are somewhat distorted, with
most of the charge missing from the (10,10) site common to both
cores. The two holes localized in the cores are
responsible for the fact that the meron-antimeron pair does not
collapse (due to Fermi statistics, it is impossible to have two holes
at the same site).   }
\end{figure}

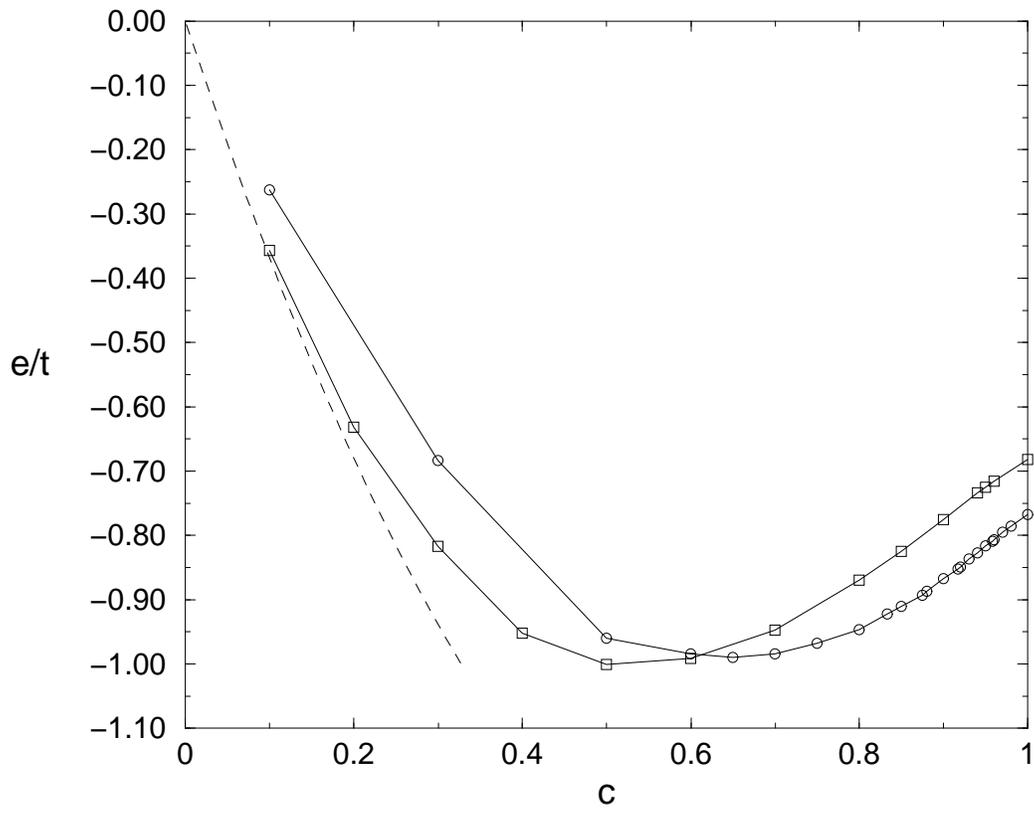
\begin{figure}
\caption{Energy per site (in units of $t$) as a function of the
electron concentration, $c=1-\delta$, for $U/t=5$.  Circles correspond
to the lowest energies found in the random trial in the spin-flux
phase ( liquid of meron-vortices), while squares correspond to the
best result of the random trial in the conventional phase (stripes).
The dashed line shows the exact value for $U=0$ (non-interacting
case). At low doping (high electron concentration) the liquid of
meron-vortices of the spin-flux phase has a lower energy than the
stripes of the conventional phases. However, at higher dopings
($\delta > 0.4$) the conventional phase becomes stable.  }
\end{figure}

\begin{figure}
\caption{Self-consistent spin distribution for the configuration of
lowest energy found at $\delta=0.08$ (8 holes), starting from an
initial random distributions, for $U/t=5$ in the spin-flux phase.
Four meron-antimeron pairs appear. We have marked with M the
plaquettes on which merons are centered, and with A the plaquettes on
which antimerons are centered.  A meron and an antimeron are ``split''
between the two opposing boundaries (we have imposed cyclic boundary
conditions).  }
\end{figure}

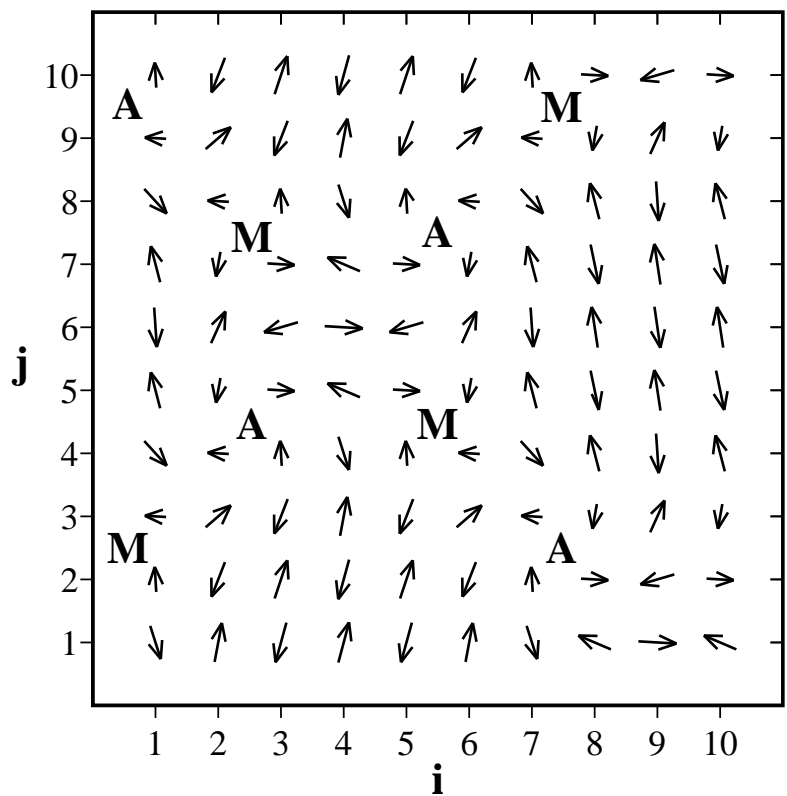
\begin{figure}
\caption{Self-consistent spin distribution for the configuration of
lowest energy found at $\delta=0.15$, starting from initial random
distributions, for $U/t=5$ in the conventional phase.  A closed
charged stripe appears (cyclic boundary conditions were imposed). The
AFM magnetic order is switched from one phase to the other one (up
$\rightarrow$ down and viceversa) as the stripe is crossed.  }
\end{figure}

\begin{figure}
\caption{Self-consistent spin distribution for the configuration of
lowest energy found at $\delta=1/8$, for $U/t=5$ in the spin-flux
phase.  An ordered crystal of charged merons and antimerons is
created.  }
\end{figure}

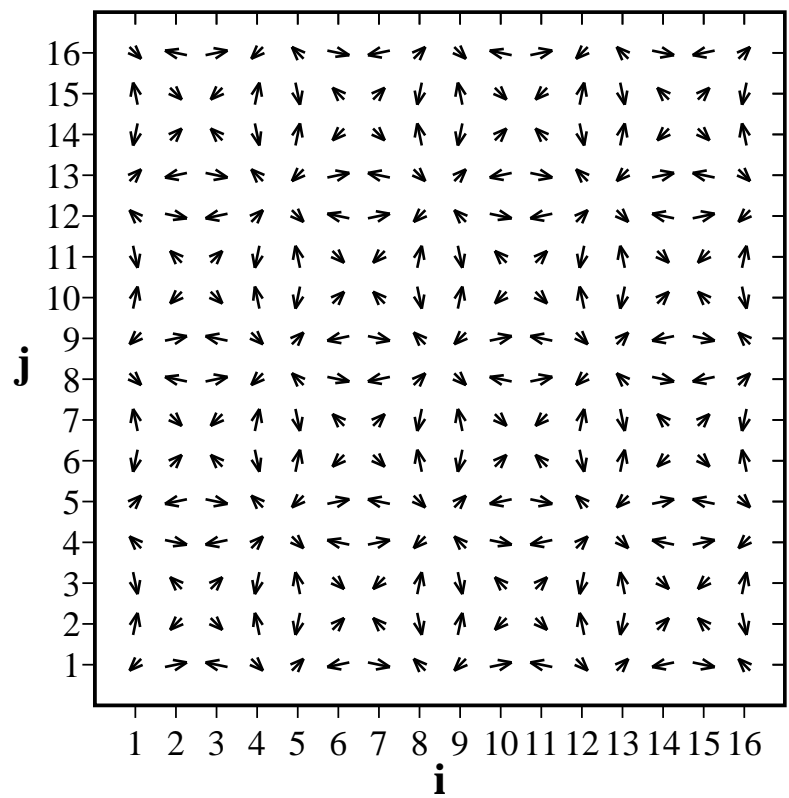
\begin{figure}
\caption{Self-consistent charge distribution for the configuration of
lowest energy found at $\delta=1/8$, for $U/t=5$ in the spin-flux phase.
An ordered crystal of charged merons and antimerons is created. }
\end{figure}

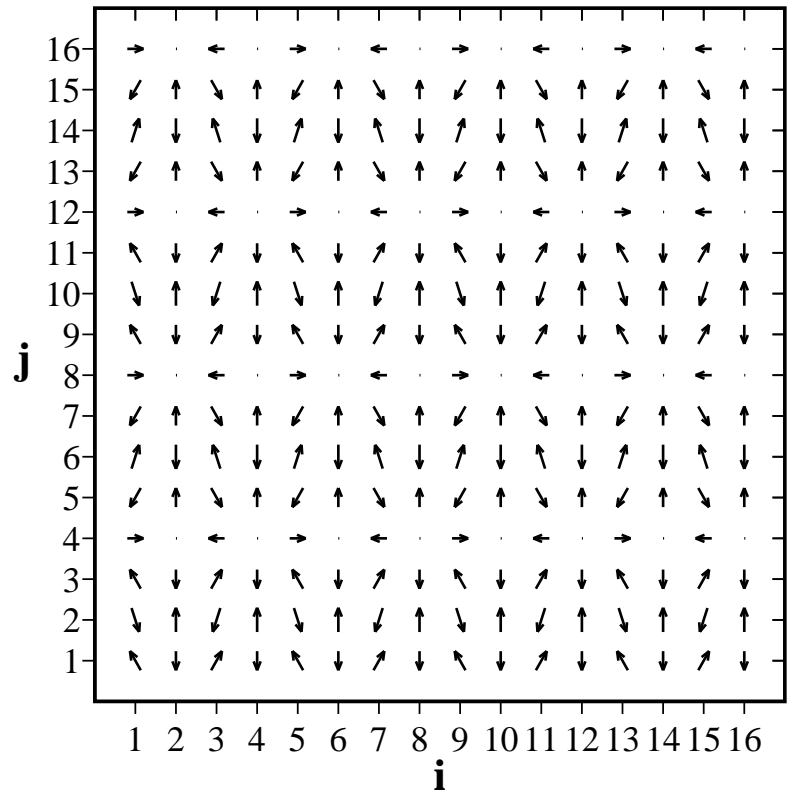
\begin{figure}
\caption{Self-consistent spin distribution for the
configuration of lowest energy found after adding a $5 \%$ anisotropy
in the hopping integral, at $\delta=1/8$, 
for $U/t=5$ in the spin-flux phase.
 The merons and antimerons rearrange on  horizontal lines, leading to
a structure similar to that suggested by Tranquada in Ref. 32.
}
\end{figure}

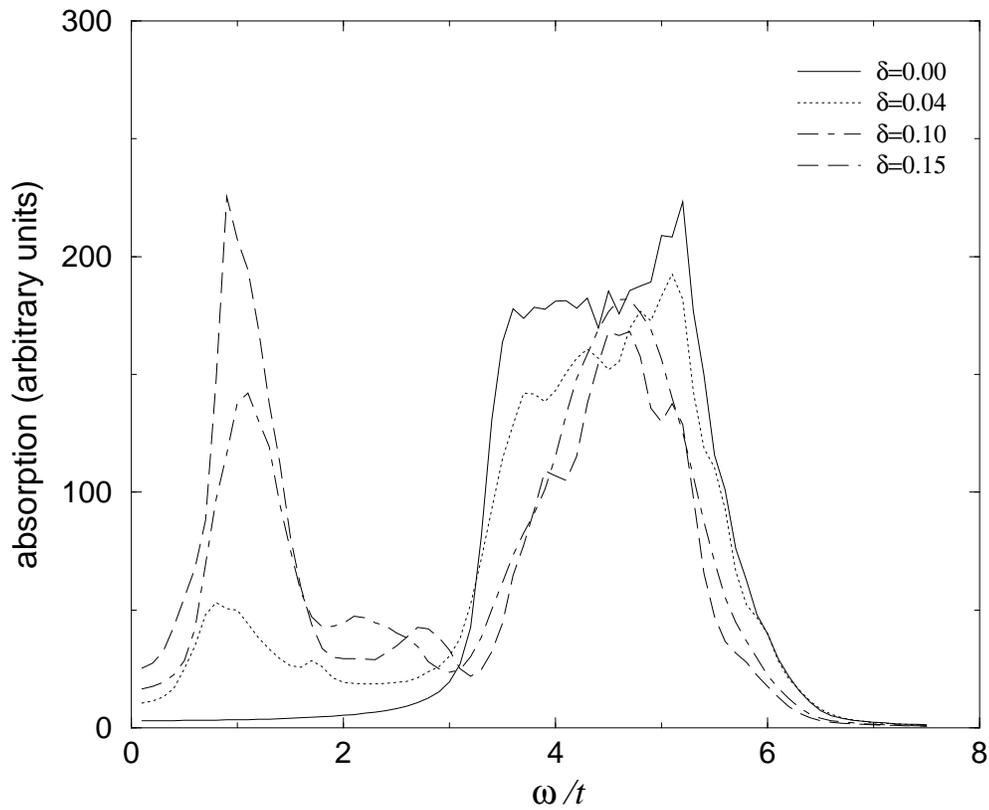
\begin{figure}
\caption{Optical absorption (arbitrary units) as a function of the
frequency (in units of $t$) for various dopings. The Hubbard parameter
is $U/t =5$, and the corresponding self-consistent lowest-energy
configurations of frozen liquids of merons and antimerons were used.
The damping coefficient is $\Gamma=0.1 t$. As the doping increases, a
broad band due electronic excitations from the lower Mott-Hubbard gap
to the discrete, empty, meron-induced levels develops deep into the
gap.  }
\end{figure}

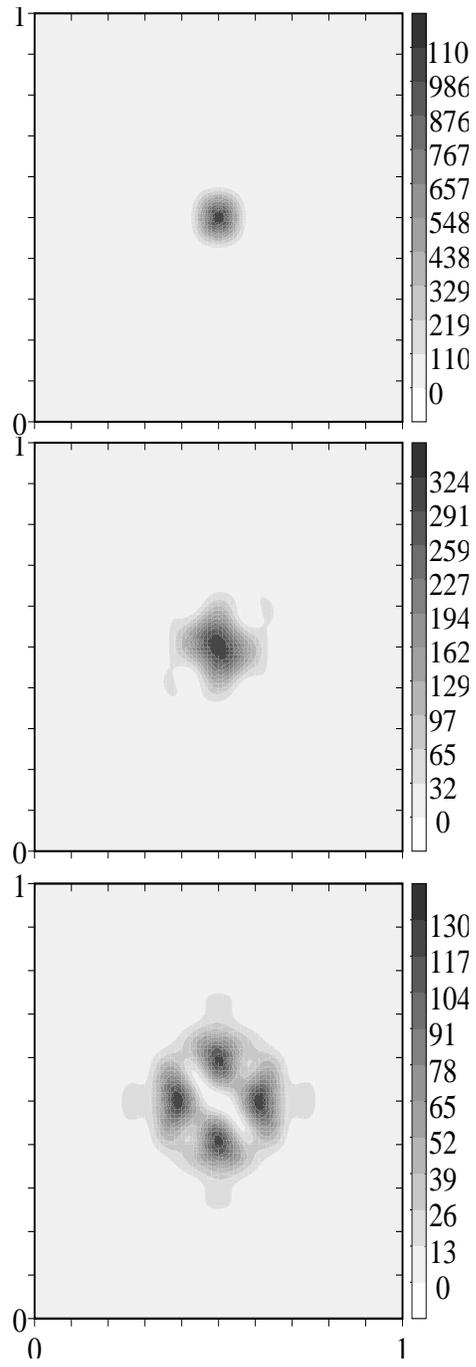
\begin{figure}
\caption{The static magnetic structure factor as a function of
$(k_x,k_y)$, measured in units of $2\pi/a$. The first picture
corresponds to $\delta=0.00$ and has the large magnetic peak at
$(\pi/a, \pi/a)$. As the doping increases to $\delta=0.05$ (second
picture) and $\delta=0.08$ (third picture) this magnetic peak splits
into four incommensurate satellites. The Hubbard parameter is $U/t
=5$, and the corresponding self-consistent lowest-energy
configurations of frozen liquids of merons and antimerons were used.
}
\end{figure}

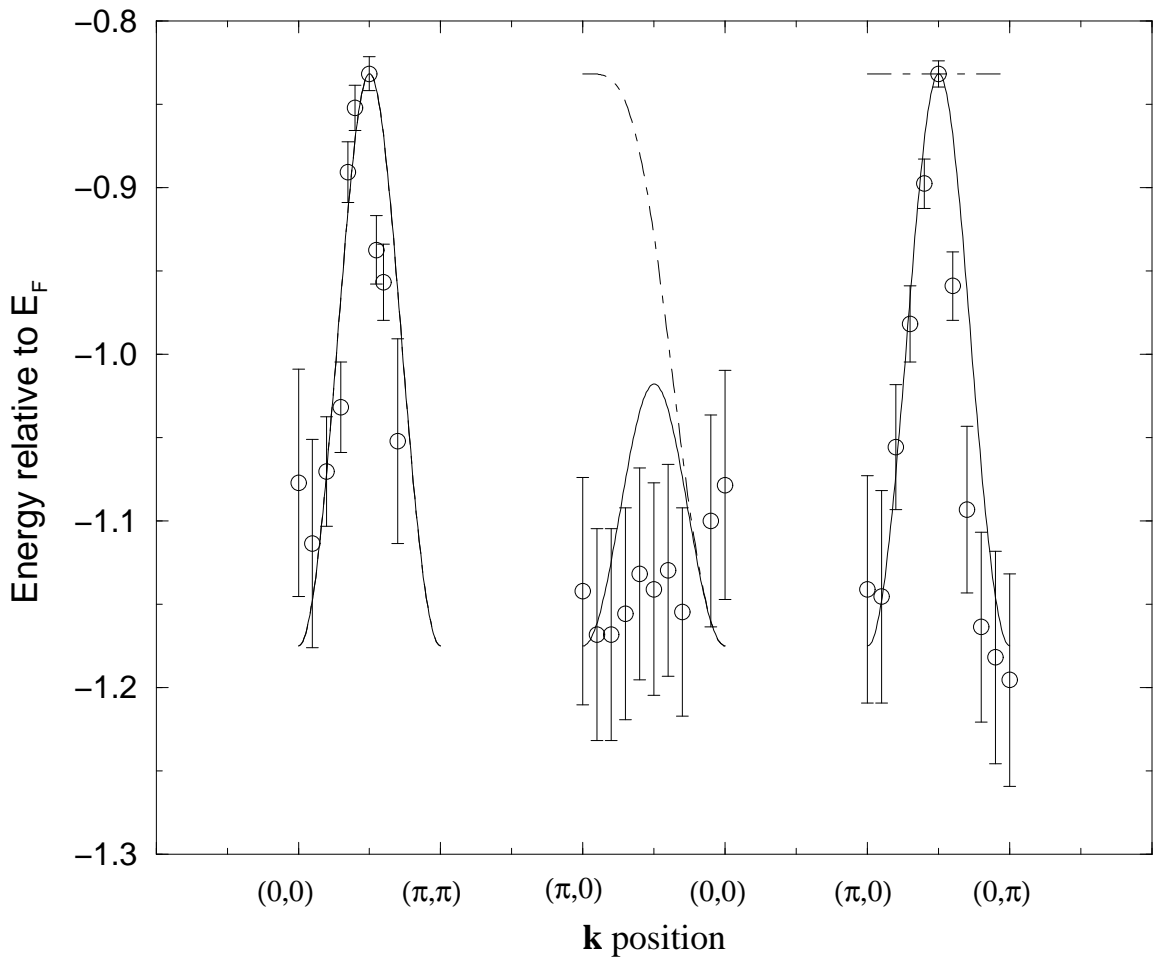
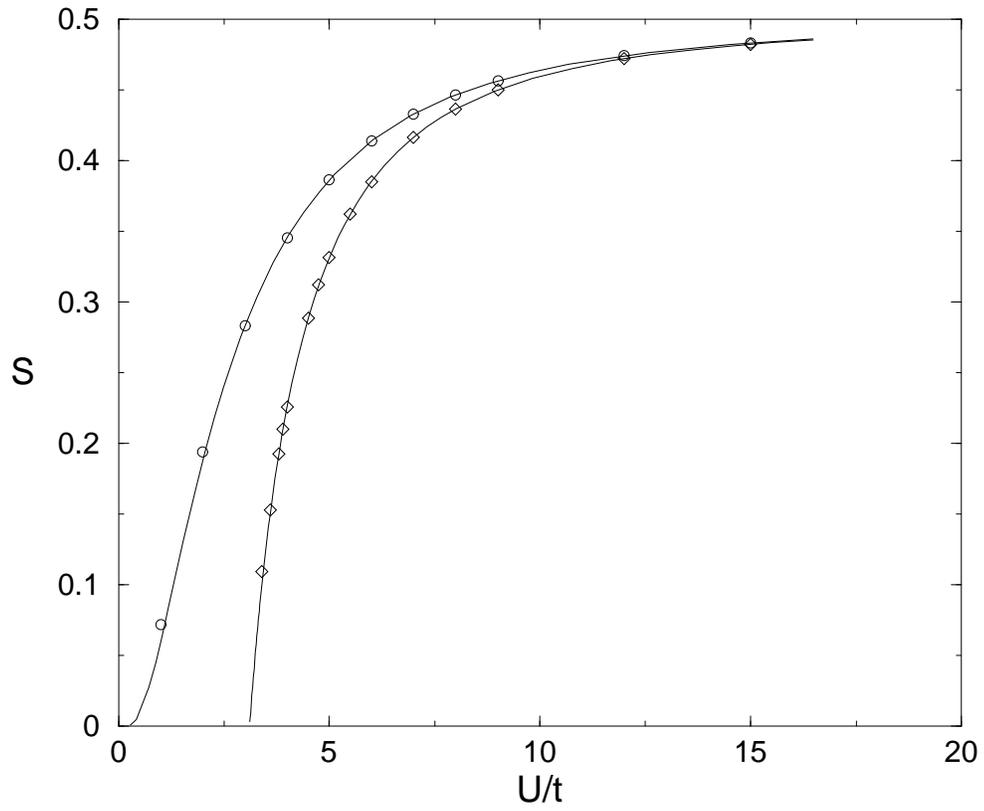
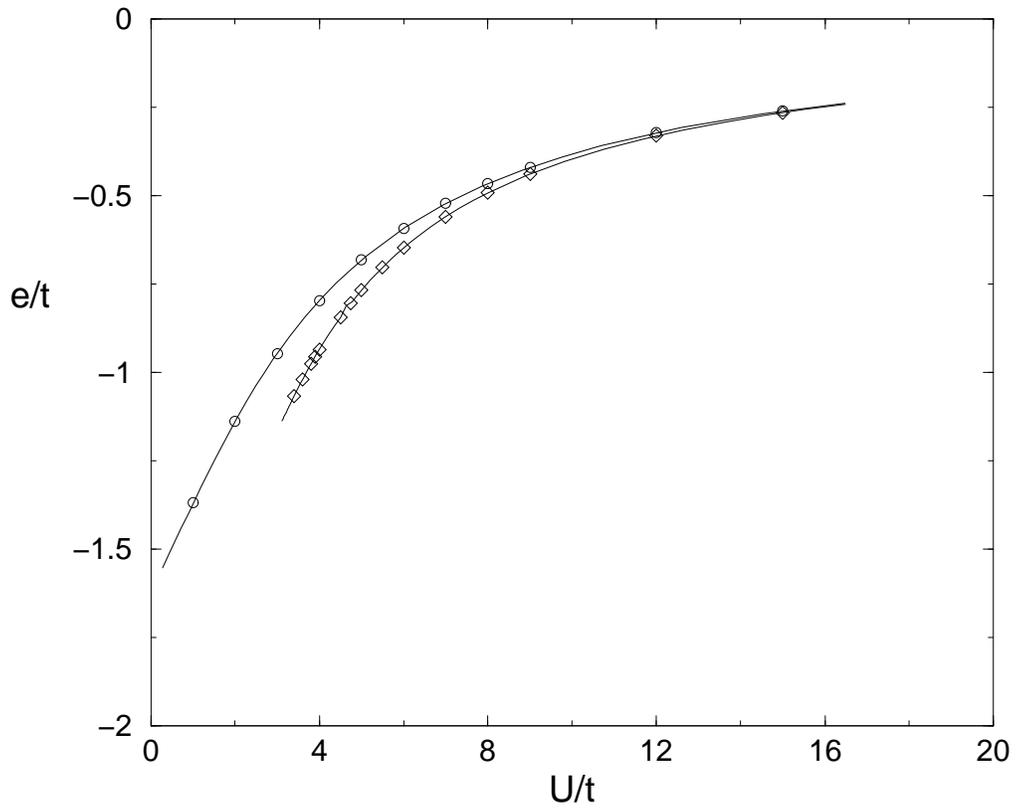
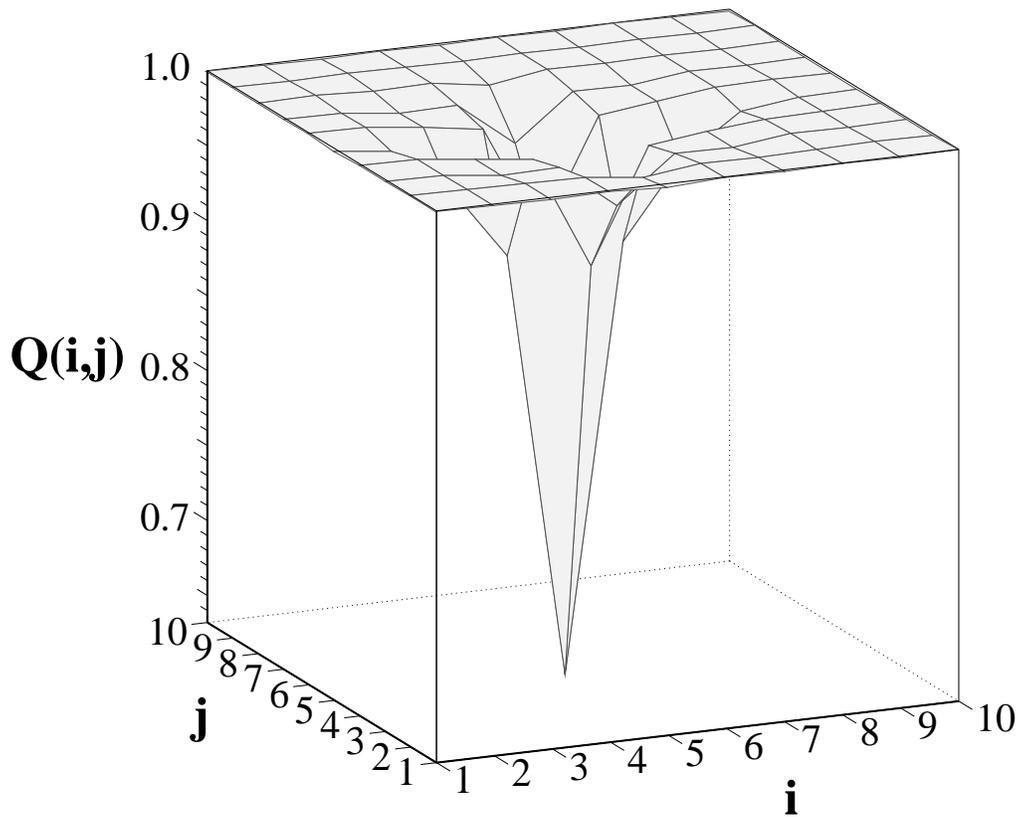
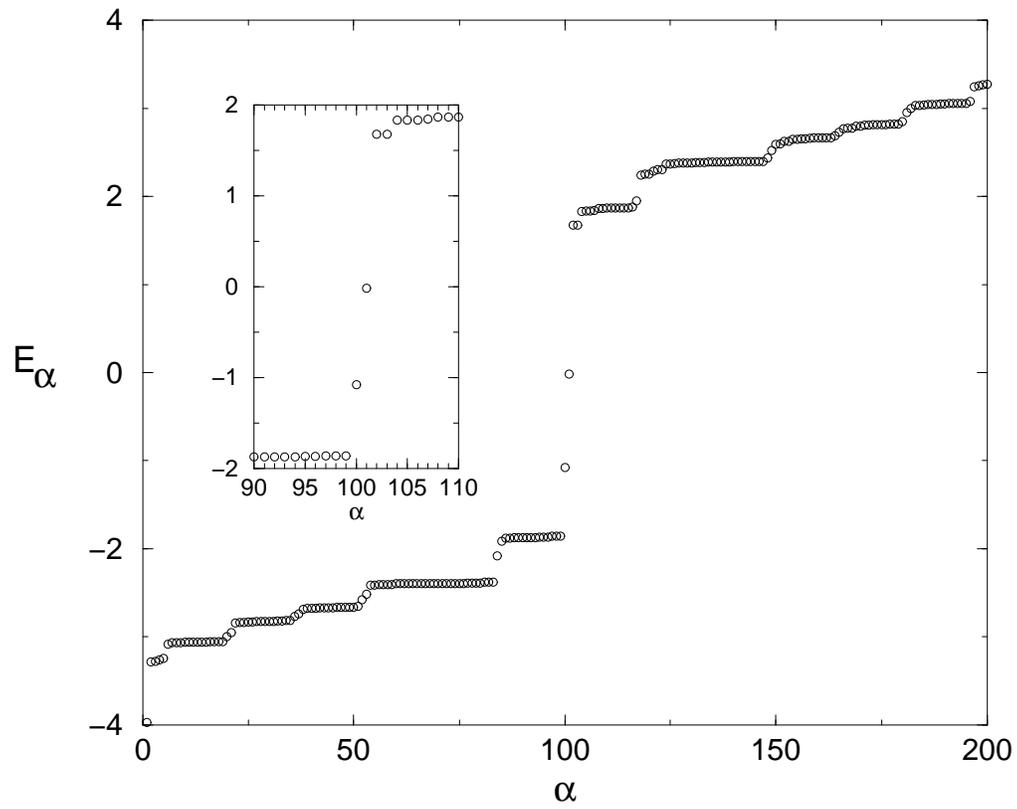
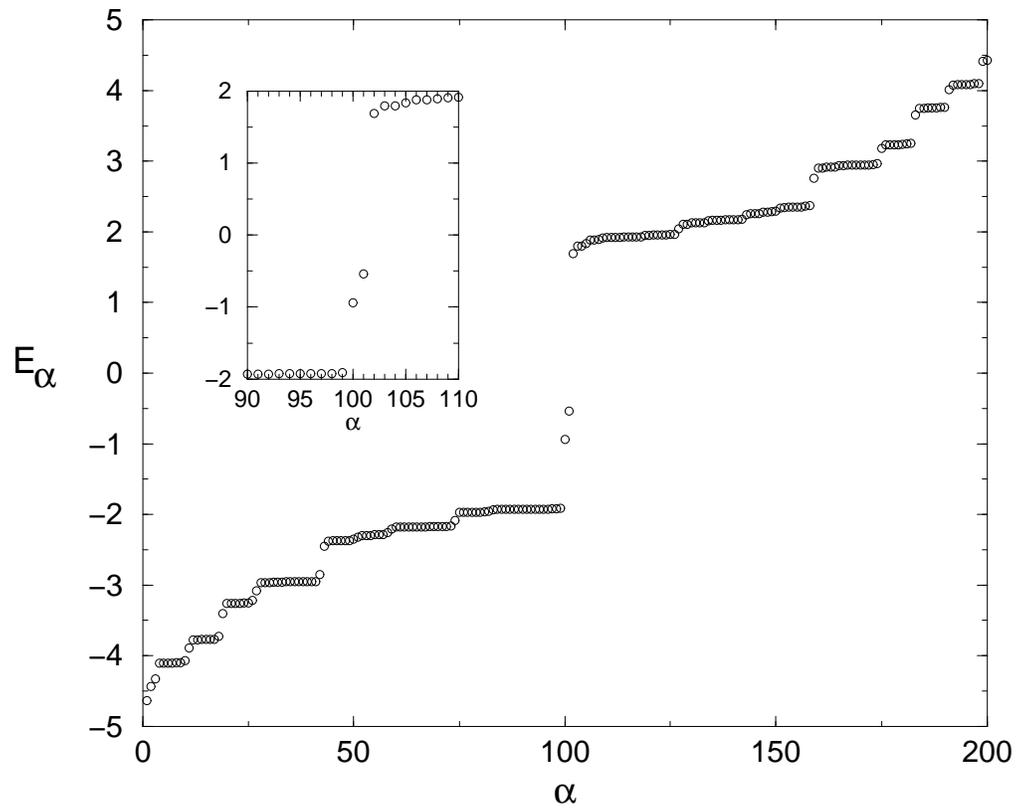
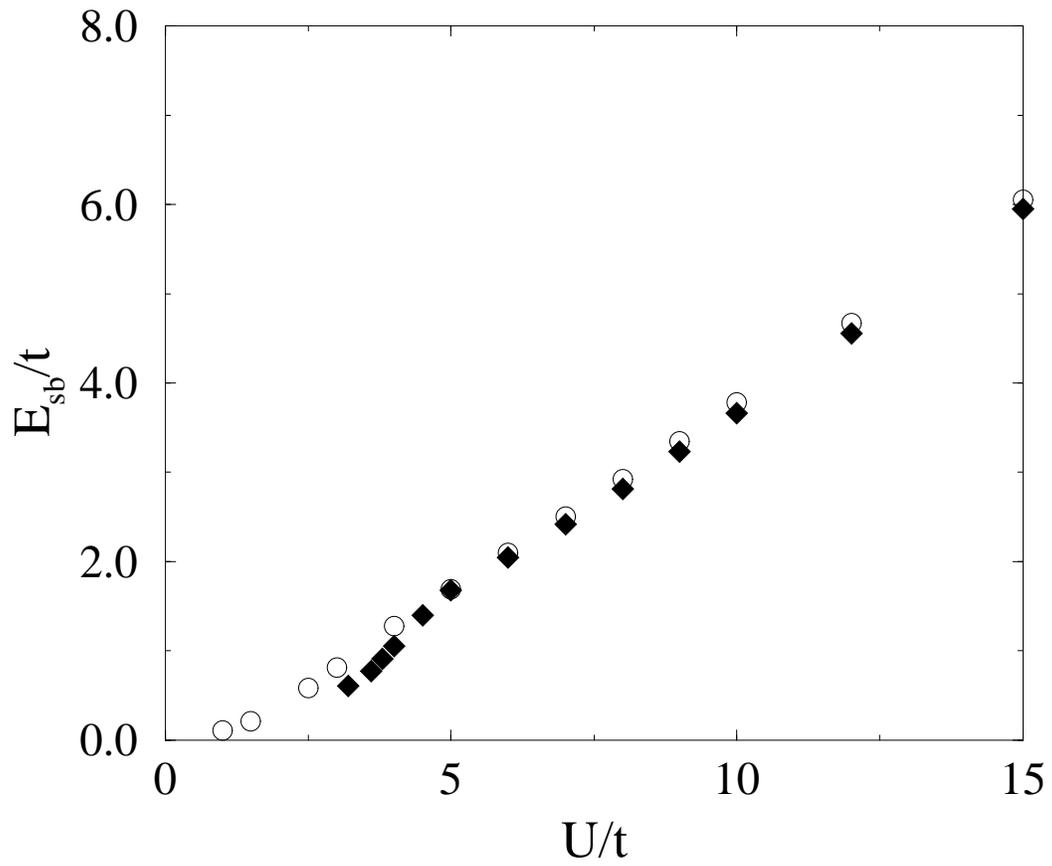
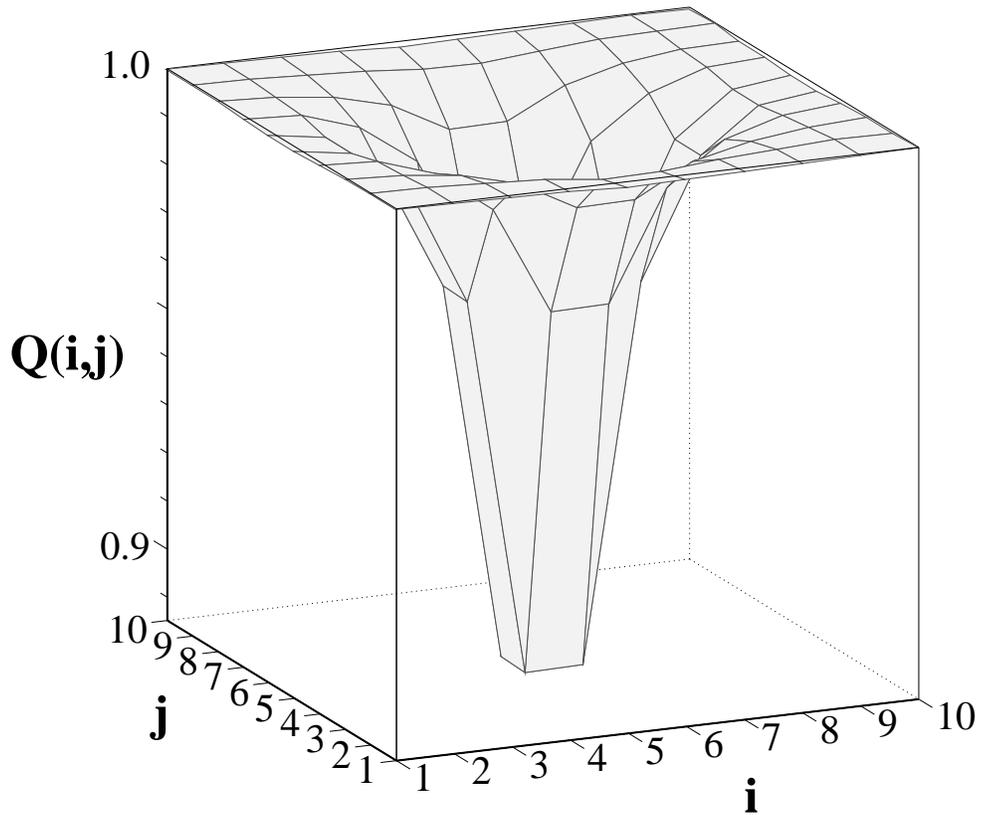
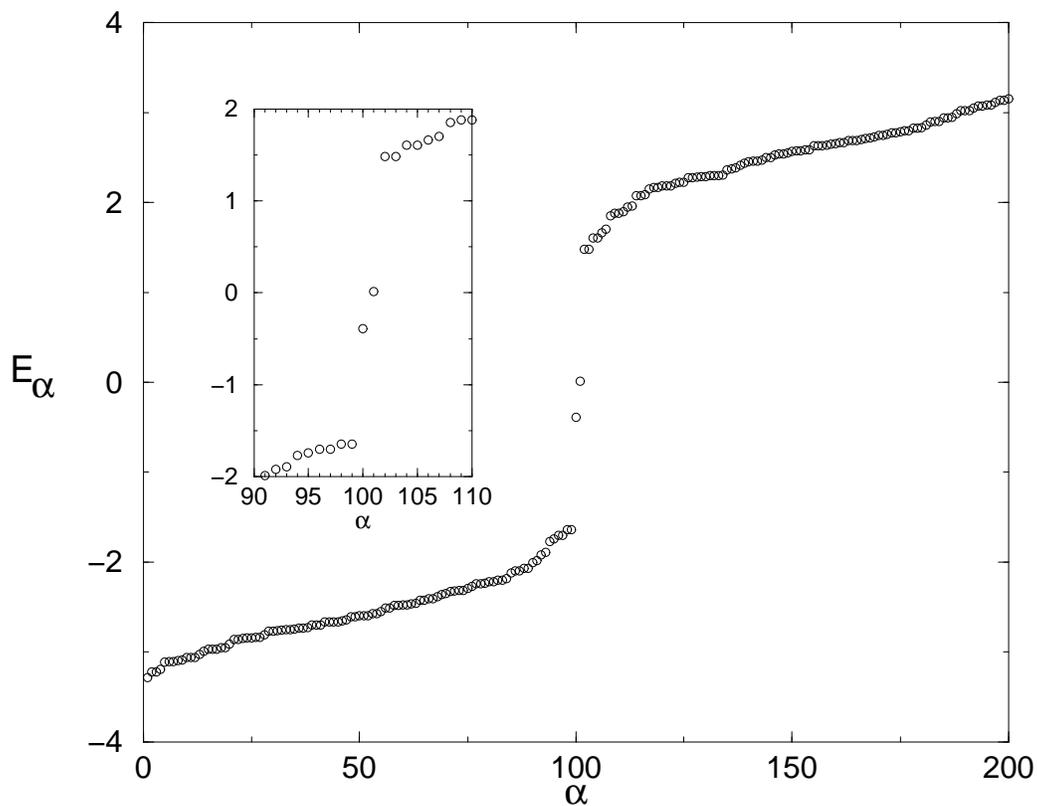
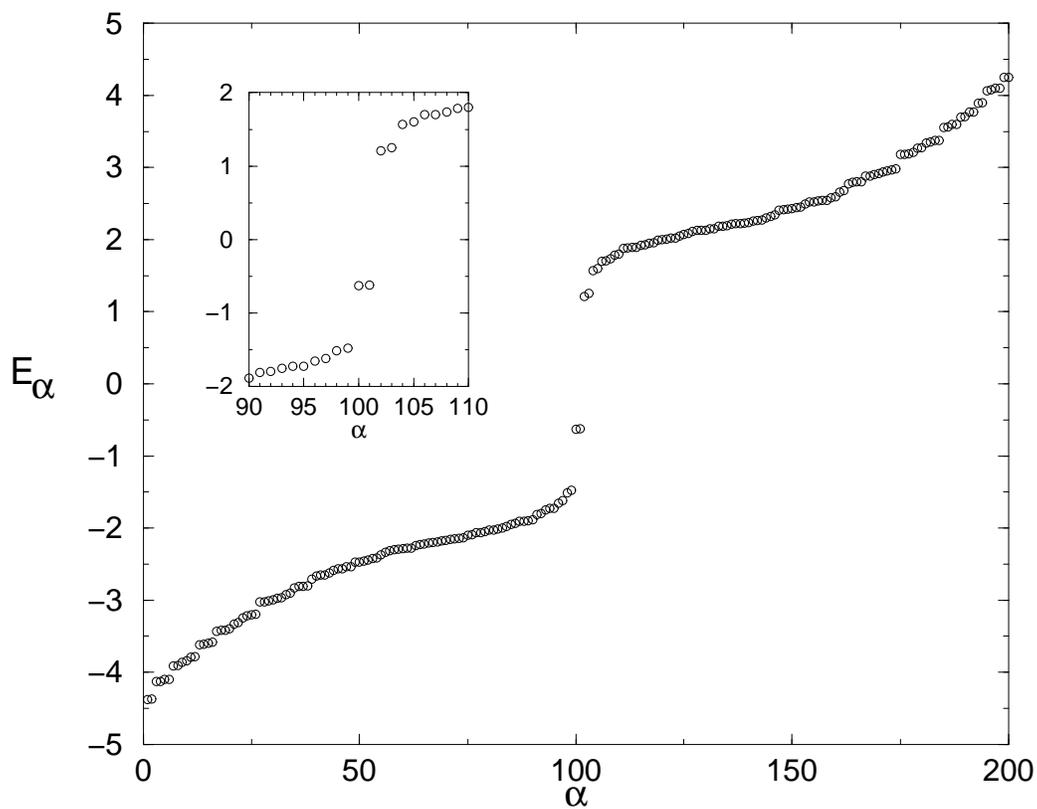
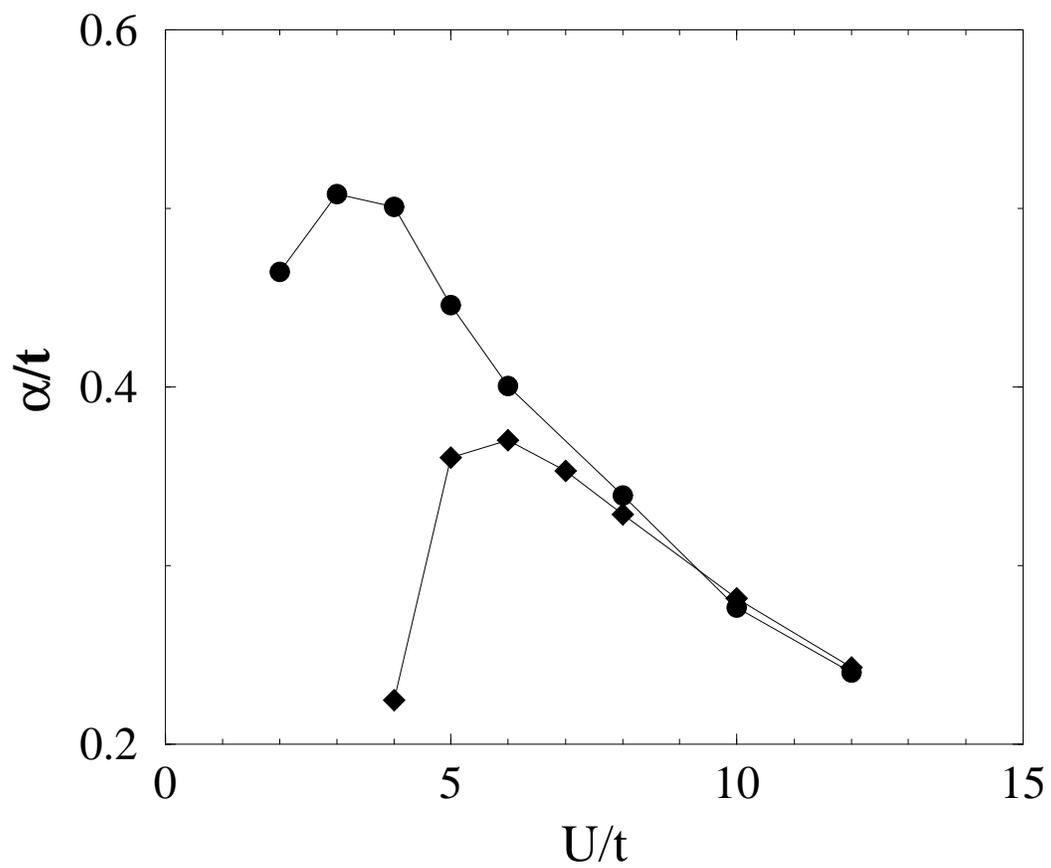
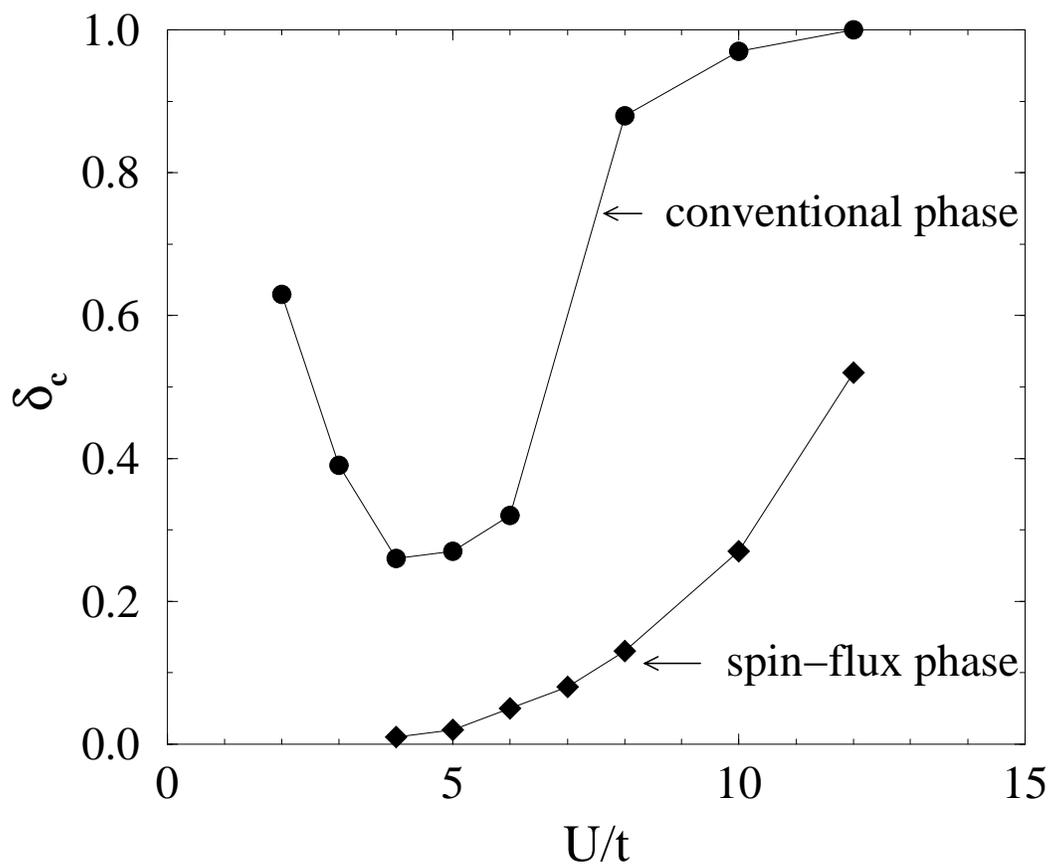
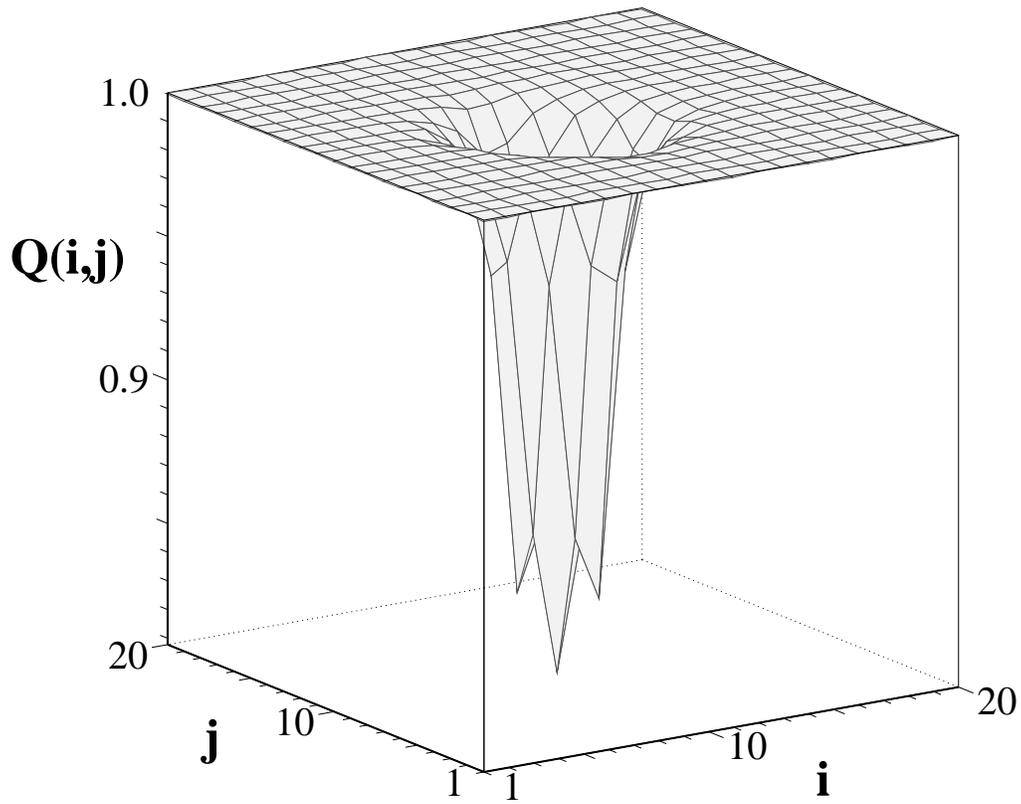
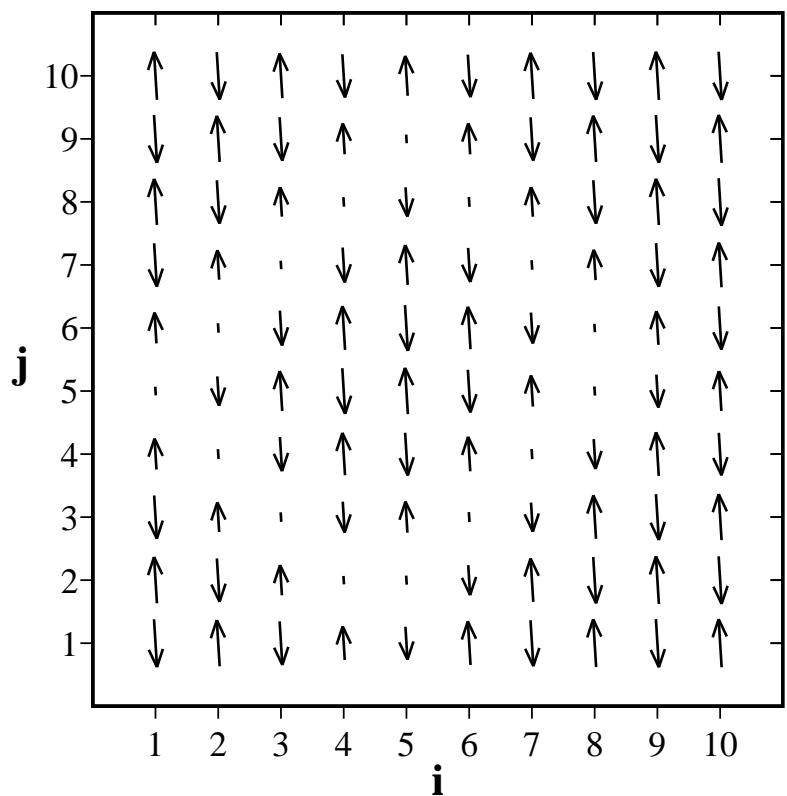
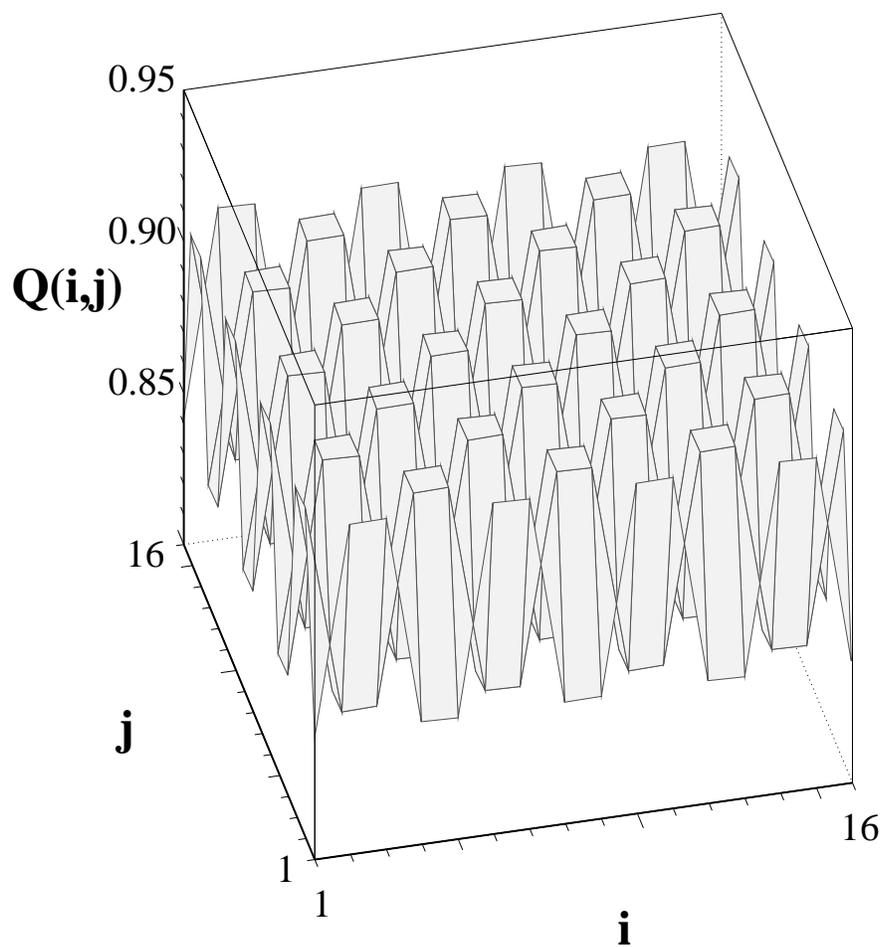
\end{document}